\documentclass[9pt]{article}   	% use "amsart" instead of "article" for AMSLaTeX format
\usepackage{geometry}                		% See geometry.pdf to learn the layout options. There are lots.
\usepackage{graphicx}				% Use pdf, png, jpg, or eps§ with pdflatex; use eps in DVI mode
\usepackage{amsmath}
\usepackage{amssymb}
\usepackage{dsfont}
\usepackage{algorithm}
\usepackage[noend]{algpseudocode}
\usepackage{hyperref}
\usepackage{doi}
\usepackage{pf2}
\usepackage{todonotes}

\newcommand{\defeq}{\overset{\mathrm{def}}{=}}

\title{GOC-Ledger: State-based Conflict-Free Replicated Ledger from Grow-Only Counters}
\author{Erick Lavoie \\
University of Basel \\
\href{mailto:erick.lavoie@unibas.ch}{erick.lavoie@unibas.ch}}

\begin{document}
\maketitle

\begin{abstract}
Conventional blockchains use consensus algorithms that totally order updates across all accounts, which is stronger than necessary to implement a replicated ledger. This makes updates slower and more expensive than necessary. More recent consensus-free replicated ledgers forego consensus algorithms, with significant increase in performance and decrease in infrastructure costs. However, current designs are based around reliable broadcast of update operations to all replicas which require reliable message delivery and reasoning over operation histories to establish convergence and safety.

In this paper, we present a replicated ledger as a state-based conflict-free replicated data type (CRDT) based on grow-only counters. This design provides two major benefits: 1) it requires a weaker eventual transitive delivery of the latest state rather than reliable broadcast of all update operations to all replicas; 2) eventual convergence and safety properties can be proven easily without having to reason over operation histories: convergence comes from the composition of grow-only counters, themselves CRDTs, and safety properties can be expressed over the state of counters, locally and globally. In addition, applications that tolerate temporary negative balances require no additional mechanisms and applications that require strictly non-negative balances can be supported by enforcing sequential updates to the same account across replicas.

Our design is sufficient when executing on replicas that might crash and recover, as common in deployments in which all replicas are managed by trusted entities. It may also provide a good foundation to explore new mechanisms for tolerating adversarial replicas.
\end{abstract}

\section{Introduction}
\label{sec:introduction}

% Consensus is not necessary
Global replicated ledgers, also known as \textit{blockchains}, provide the ability for untrusted parties to exchange tokens world-wide, with Bitcoin~\cite{nakamoto2008bitcoin} and Ethereum~\cite{buterin2014next} being the most popular examples. It has been shown that using consensus algorithms to totally order updates across all accounts in replicated ledgers is stronger than necessary to prevent overspending~\cite{guerraoui2021consensus}: sequentially ordering updates on the same account (\textit{e.g.},~\cite{frey:hal-03346756}) and even only the outgoing transactions is sufficient  (\textit{e.g.},~\cite{collins2020broadcast-payment}). Foregoing the use of consensus algorithms drastically increases throughput and decreases latency~\cite{collins2020broadcast-payment,baudet2020fastpay}. Published consensus-free replicated ledgers~\cite{collins2020broadcast-payment,baudet2020fastpay,sliwinski2020abc,guerraoui2021consensus,auvolat2021money,frey:hal-03346756,kuznetsov2021permissionless,cholvi2021bdso,georghiades2021needs} are based on broadcast abstractions that require reliable delivery of account operations. 

% Opportunity of state-based CRDTs
In parallel to work on replicated ledgers, Conflict-Free Replicated Data Types (CRDTs)~\cite{shapiro:hal-00932836} have been proposed to simplify the design and implementation of distributed systems. CRDTs are replicated mutable objects designed according to simple constraints to ensure convergence to the same state \textit{eventually}, \textit{i.e.}, at some point in the future after updates have stopped, and \textit{automatically}, \textit{i.e.} using deterministic conflict-resolution rules in the presence of concurrent updates. The state-based declination of CRDTs~\cite{shapiro:hal-00932836} requires only 1) eventual transitive delivery of the latest state rather than reliable broadcast of all update operations to all replicas, and 2) a simple merging procedure between different states. The state-based approach tolerates by design lost, re-ordered, stale, and duplicate updates containing copies of other replicas' state by ensuring the possible states and operations follow a few simple constraints.

In this paper, \textit{we show that a replicated ledger can be implemented as a state-based conflict-free data type based on the composition of grow-only counters}, which we call \textit{Grow-Only-Counters-Ledger} (GOC-Ledger). The key idea is to use a separate grow-only counter for every kind of state-updating operation, which represents the total contribution of all past similar operations to the current balance. For example, a specific grow-only counter is used to track the total amount ever sent  from a sender account to a specific receiver account. The current balance of an account is then computed by adding every balance-increasing counters and subtracting every balance-reducing counters. 

To the best of our knowledge, we are the first to propose this design, which provides numerous benefits. First, as for all state-based CRDTs, it lowers the requirements on the communication channels and connection topology between replicas. Effectively, grow-only counters act as a summary of all previous operations of the same kind, which removes the need to reason about and individually track previous operations. Second, eventual convergence of all account replicas to the same balance can be simply proven by the composition of grow-only counters, that are themselves simple state-based CRDTs. The convergence of account replica balances is guaranteed even in the presence of overspending, which we show can only occur in the presence of concurrent updates to the same account from multiple replicas.

% Local crypto-tokens opportunity
%The use of crypto-tokens has been popularized by global replicated ledgers, also known as \textit{blockchains}, for anonymous transactions between untrusted parties world-wide, with Bitcoin~\cite{nakamoto2008bitcoin} and Ethereum~\cite{buterin2014next} being the most popular.
%However, high transactions fees preclude their use for many local economic applications~\cite{lavoie2022localcryptotokens}.
%Local economies are characterized by repeated interactions within a community of participants that know each other, providing higher trust levels than is otherwise assumed by participants on global blockchains. Revisit underlying economic assumptions behind the current designs to enable simpler and more affordable infrastructure in local contexts.

As additional benefits, a GOC-Ledger might be particularly appropriate for supporting local economic applications~\cite{lavoie2022localcryptotokens}, that are characterized by repeated interactions within a community of participants that know each other. Some local applications may tolerate temporary negative balances, for example because existing trust between participants or reputation loss are sufficient to incentivize maintaining a non-negative balance. For other applications in which a non-negative balance should be maintained at all times, our design can be combined with an additional mechanism that ensures all updates to the same account are ordered, similar to existing designs~\cite{collins2020broadcast-payment,baudet2020fastpay}. Either case should be relatively straight-forward to implement when all replicas are managed by a single organization or a restricted set of mutually-trusting entities.

However, our design is not sufficient by itself to operate in the presence of adversarial replicas, as happens in peer-to-peer settings. It is still an open research question to design practical and frugal replicated ledgers that work in those settings. We believe GOC-Ledger is a good foundation for such explorations, and plan to use it to design the necessary mechanisms. 

In the rest of this paper,  we present the design of the GOC-Ledger (Section~\ref{sec:design}), we prove its convergence, as well as its balance safety and liveness properties (Section~\ref{sec:proofs}), we position it in the context of other related work  and highlight open challenges to tolerate adversarial failures (Section~\ref{sec:related-work}). We finally conclude with a brief summary and outline future research directions (Section~\ref{sec:conclusion}).

\newpage

\section{Grow-Only-Counters (GOC) Ledger}
\label{sec:design}

In this section, we present the design of a GOC-Ledger in two parts: we first present state-based conflict-free replicated \textit{accounts} (Section~\ref{sec:account}) and then the \textit{ledger} proper (Section~\ref{sec:ledger}). Our notation and conventions are explained as they are introduced and collected in Appendix~\ref{apdx:notation}.

\subsection{Account}
\label{sec:account}

Account replicas track the current balance of tokens for a given owner identified by their \textit{identifier} ($id$). The state of an account is decomposed into multiple internal counters such that all counters are \textit{monotonically increasing}, \textit{i.e.}, every operation leaves counters either unchanged or larger. 

Before explaining what each internal counter represents, we give an overview of account operations. Each owner can  \textit{create} new tokens, if they have the capability; \textit{burn} tokens they own; and transfer some or all their tokens to other accounts, identified by different $id$s. We split the transfer into two operations: the sender \textit{gives} tokens to a receiver account, which immediately decreases the balance of the sender; however, the balance of the receiver is only increased after an explicit \textit{acknowledgement}. This decomposition enables the reception of tokens to be tracked as a separate event: this allows, for example, other replicas to check if the receiver account has received (and accepted) the tokens.\footnote{It is possible to always automatically acknowledge the reception of tokens immediately upon reception and not expose the acknowledgment operation to the user. However, having the operation performed explicitly also gives the opportunity to the receiver to ignore tokens sent to them. We therefore prefer the explicit operation.} The implementation follows easily, with each operation modifying separate internal counter(s). 

We first present the state and operations in the following section (Section~\ref{sec:account:state-operations}), and the ordering and merging operations necessary for concurrency, in the next (Section~\ref{sec:account:ordering-merging}).

\subsubsection{State and Operations}
\label{sec:account:state-operations}

The implementation of the state and corresponding operations is listed in Algorithm~\ref{alg:account} and ~\ref{alg:query}. We now present each operation in turn.

\begin{algorithm}
\begin{algorithmic}[1]
   \State \textbf{Require} $\mathds{C}$, the set of identifiers allowed to create tokens
   \State
   \Function{initialize$_\mathds{A}$}{\textit{id}}
    	\State $A_{\scriptsize\textbf{id}} \leftarrow id$
	\State $A_\uparrow ~\leftarrow 0$       \Comment{Created amount (real positive number)}
	\State $A_\downarrow ~\leftarrow 0$   \Comment{Burned amount (real positive number)}
	\State $A_\rightarrow \leftarrow \{ \}$ \Comment{GivenTo: Dictionary of $\textit{id}$s associated to $\textit{amount}$s}
	\State $A_\leftarrow \leftarrow \{ \}$ \Comment{AckFrom: Dictionary of $\textit{id}$s associated to  $\textit{amount}$s}
    	\State \Return $A$
    \EndFunction
    \State
    \Function{create}{$A,\textit{amount}$} 
    	\If{$A_{\scriptsize\textbf{id}} \in \mathds{C} \wedge \textit{amount} > 0$} 
	      	 \State $A' \leftarrow \textit{copy}(A)$
		 \State $A'_\uparrow \leftarrow A_\uparrow + \textit{amount}$
		 \State \Return $A'$
	\Else
		\State \Return $A$
	\EndIf

    \EndFunction
    \State
     \Function{burn}{$A,\textit{amount}$}  
    	\If{$\textit{amount} > 0 \wedge \texttt{balance}(A) \geq \textit{amount}$} \label{ln:burn-balance-check}
		 \State $A' \leftarrow \textit{copy}(A)$
		 \State $A'_\downarrow \leftarrow A_\downarrow + \textit{amount}$
		 \State \Return $A'$
	\Else
		\State \Return $A$
	\EndIf
    \EndFunction
    \State
    \Function{giveTo}{$A,\textit{amount}, \textit{id}$}  
    	\If{$\textit{amount} > 0 \wedge \texttt{balance}(A) \geq \textit{amount}$}  \label{ln:give-balance-check}
	        \State $A' \leftarrow \textit{copy}(A)$
		\If{$id \notin A_{\rightarrow *}$}
			\State $A'_{\rightarrow}[id] \leftarrow \textit{amount}$ 
		\Else
		         \State $A'_{\rightarrow}[id] \leftarrow A_{\rightarrow}[id] + \textit{amount}$ 
		\EndIf
		\State \Return $A'$
	\Else
		\State \Return $A$
	\EndIf

    \EndFunction
    \State
    \Function{ackFrom}{$A, B$}
        \If{$\texttt{unackedFrom}(A,B) > 0$}
                \State $A' \leftarrow \textit{copy}(A)$
		\If{$B_{\scriptsize\textbf{id}} \notin A_{\leftarrow *}$}
			 \State $A'_{\leftarrow}[B_{\scriptsize\textbf{id}}] \leftarrow B_{\rightarrow}[A_{\scriptsize\textbf{id}}]$
		\Else
			 \State $A'_{\leftarrow}[B_{\scriptsize\textbf{id}}] \leftarrow \texttt{max}(A_{\leftarrow}[B_{\scriptsize\textbf{id}}], B_{\rightarrow}[A_{\scriptsize\textbf{id}}])$ 
		\EndIf
		\State \Return $A'$
        \Else
        		\State \Return $A$
        \EndIf

    \EndFunction
    \end{algorithmic}
\caption{\label{alg:account} Account: State Initialization and State-Changing Operations}
\end{algorithm}

\begin{algorithm}
\begin{algorithmic}[1]
    \State
        \Function{balance}{A}
        \State $\textit{debits} \leftarrow A_\uparrow + \sum\limits_{id \in A_{\leftarrow *}} A_{\leftarrow}[id]$ 
        \Comment{Keys of $A_\leftarrow$ written $A_{\leftarrow *}$}
        \State $\textit{credits} \leftarrow A_\downarrow + \sum\limits_{id \in A_{\rightarrow *}} A_{\rightarrow}[id]$
        \Comment{Keys of $A_\rightarrow$ written $A_{\rightarrow *}$}
        \State \Return $\textit{debits} - \textit{credits}$
    \EndFunction
    \State
    \Function{unackedFrom}{$A, B$}
        \If{$A_{\scriptsize\textbf{id}} \in B_{\rightarrow *} \wedge B_{\scriptsize\textbf{id}} \in A_{\leftarrow *} $}
        		\State \Return $B_{\rightarrow A_{\tiny\textbf{id}}} - A_{\leftarrow B_{\tiny\textbf{id}}}$
	\ElsIf{$A_{\scriptsize\textbf{id}} \in B_{\rightarrow *}$}
		\State \Return $B_{\rightarrow A_{\tiny\textbf{id}}}$
	\Else
		\State \Return $0$
	\EndIf
    \EndFunction
    \end{algorithmic}
\caption{\label{alg:query} Account: Query Operations}
\end{algorithm}

A replica of an account belonging to owner $id$ is created with \texttt{initialize} (Alg.~\ref{alg:account}). This initializes the internal identifier $A_{\textbf{id}}$ to $id$, the number of created tokens $A_\uparrow$ to $0$, the number of burned tokens $A_\downarrow$ to $0$, the dictionary of given tokens $A_\rightarrow$ to an empty dictionary, and the dictionary of acknowledged tokens $A_\leftarrow$ to an empty dictionary as well.\footnote{The direction of the arrow notation suggests the corresponding direction of funds: up or down for creation or destruction of tokens, and  left or right for tokens flowing into and out of an account.}

The \texttt{create} operation (Alg.~\ref{alg:account}) takes the current state of an account $A$ and a requested \textit{amount} then returns a new state $A'$.\footnote{A practical implementation may actually encapsulate the state of an account in an object and modify it in place. We explicitly associate the different states to different variables to make the labels of the states in proofs (Section~\ref{sec:proofs}) easier to relate to the implementation.} Only allowed creators, \textit{i.e.}, those with ids within the set of allowed identifiers $\mathds{C}$, may create new tokens.\footnote{The set of allowed creator identifiers might be distributed to replicas beforehand or derived from cryptographic mechanisms. The design of such mechanisms is outside the scope of this paper.} If the owner of the account is indeed allowed to create tokens, the creation counter is increased by \textit{amount} and $A'_\uparrow = A_\uparrow + \textit{amount}$. This has the effect of increasing the balance by \textit{amount}. Otherwise, the counter is unchanged ($A'_\uparrow=A_\uparrow$) and the balance remains the same.

The \texttt{burn} operation (Alg.~\ref{alg:account}) is similar to \texttt{create} with an opposite effect on the balance: it takes the current state of an account $A$ and a requested \textit{amount} to burn, then returns a new state $A'$. If the balance is equal or larger than \textit{amount}, the burn counter is increased by \textit{amount}, resulting in $A'_\downarrow = A'_\downarrow + \textit{amount}$, and the balance correspondingly decreased by the same amount resulting in a non-negative balance. Otherwise, $A'_\downarrow = A_\downarrow$ and the balance is unchanged. In contrast to \texttt{create}, any account is allowed to burn tokens.\footnote{Restricting burning tokens to a set of identifiers would also be possible. However, the restriction would not be effective unless transfer operations were also restricted to only valid addresses because giving to an invalid or unused address effectively makes the tokens unavailable, similar to burning. Restricting transfers to only valid addresses would require tracking the set of valid accounts and most likely synchronization with highly-available replicas. We therefore take the simpler and more open approach of allowing any account to burn tokens.}

The \texttt{giveTo} operation (Alg.~\ref{alg:account}) sends \textit{amount} tokens from account $A$ to the recipient account associated to $id$, resulting in a new state $A'$. The account state tracks the total amount ever sent to $id$ using the dictionary $A_{\rightarrow}$, using $id$ as a key to access the associated monotonic counter, written $A_{\rightarrow}[id]$. This counter is only increased if the local balance is higher or equal than the \textit{amount} sent, resulting in a non-negative balance for $A'$.

\texttt{Burn} and \texttt{giveTo} are the only two operations on accounts that may decrease the balance. As we prove later by simple induction (Section~\ref{section:proof:sequential-non-negative}), it is sufficient to sequentially order all account operations, potentially across replicas, to guarantee that the balance for that account will always be non-negative. A contrario, as we also discuss next  (Section~\ref{sec:account:ordering-merging}), concurrent updates to the same account in different replicas may eventually result in a negative balance, even if the balance resulting from individual operations was always non-negative.

The \texttt{ackFrom} operation (Alg.~\ref{alg:account}) acknowledges the reception of tokens from account $B$ on account $A$. If there are unacknowledged tokens that $B$ sent to $A$,
then $A$'s counter $A_{\leftarrow}[B_{\scriptsize \textbf{id}}]$, which tracks the total amount ever acknowledged by $A$ from $B$, is increased by the number of unacknowledged tokens, with a corresponding increase the balance of $A'$. If $A$ had never acknowledged tokens from $B$ before, the new counter value  $A'_{\leftarrow}[B_{\scriptsize \textbf{id}}]$ is simply the total amount ever sent by $B$ to $A$ up to that point, written $B_{\rightarrow}[A_{\scriptsize \textbf{id}}]$. Otherwise, if $A$ had already acknowledged tokens from $B$, the new counter value  $A'_{\leftarrow}[B_{\scriptsize \textbf{id}}]$ is the maximum of the previous value $A_{\leftarrow}[B_{\scriptsize \textbf{id}}]$ and $B_{\rightarrow}[A_{\scriptsize \textbf{id}}]$. This definition correctly handles acknowledging with a stale state  $B^{-1}$ because the corresponding counter $B^{-1}_{\rightarrow}[A_{\scriptsize \textbf{id}}]$ would be equal or smaller than the counter in the most recent state of $B$, $B_{\rightarrow}[A_{\scriptsize \textbf{id}}]$, and therefore $A$ would not be updated.\footnote{It could also be useful to check that the balance of $B$ was non-negative prior to updating the state of $A$: this would limit the propagation of overspent tokens by $B$\footnote{ This would not completely prevent the propagation because different accounts may have individually acknowledged tokens from B before B's state would reflect the overspending.}. Checking for a non-negative balance from $B$ is however superfluous if the application is actually intended to handle negative balances (Section~\ref{sec:economics})}.

As follows from the previous explanations, the \texttt{balance} of an account $A$ (Alg.~\ref{alg:query}) is the sum of counters that increase the balance (\textit{debits}) subtracted by the sum of counters that decrease it (\textit{credits}). The \texttt{balance} operation does not modify the state of the account. The debits are the sum of created tokens and the sum of the tokens acknowledged from other accounts. The credits are the sum of tokens burnt and given to other accounts.

The last operation, \texttt{unackedFrom} (Alg.~\ref{alg:query})  returns the number of tokens sent from account $B$ that have not yet been acknowledged by account $A$. It does not modify the state of $A$ nor $B$. This is implemented by subtracting the tokens acknowledged by $A$ from $B$, from the tokens sent by $B$ to $A$. If the result is greater than zero, there are unacknowledged tokens.

\subsubsection{Comparing and Merging Concurrent States}
\label{sec:account:ordering-merging}

We now augment our previous account definition with one relation and one operation to enable comparing states and merging concurrently modified states. The augmented definition forms a \textit{monotonic join semi-lattice}, which informally, is a mathematical structure that combines an ordering between states with constraints on operations that captures the notion of progress without having to explicitly reason about time and ordering between different operations. We now present its different components.

The relation $\leq_\mathds{A}$ (Alg.~\ref{alg:account-ordering}) establishes an ordering between states of the same account, \textit{i.e.}, account states with the same id. The operations of the previous section (Alg.~\ref{alg:account} and ~\ref{alg:query}) have been carefully designed to be \textit{monotonic}, \textit{i.e.} they either do not modify the state of an account, or result in a new state $A'$ larger than the previous state $A$ according to $\leq_\mathds{A}$. Because of the monotonicity of the operations, if $A <_\mathds{A} A'$,\footnote{$A \leq_\mathds{A} A'$ and $A \neq A'$} then $A$ has happened before $A'$ and therefore $A'$ also contains all the updates that led to $A$. This ordering is however \textit{partial}: \textit{i.e.} if neither $A <_\mathds{A} A'$ nor $A' <_\mathds{A} A$, then neither $A$ or $A'$ has happened before the other and both are therefore the result of concurrent updates. The relation $A \leq_\mathds{A} A'$ is implemented as follows: it is true if and only if the set of identifiers for given and acknowledged tokens of $A$ are subsets of those of $A'$ \textit{and} every counter of $A$ are smaller or equal to the corresponding counters in $A'$. Note that counters in $A'$ not present in $A$ have no effect, because they would be the result of operations that have happened after $A$. 

\begin{algorithm}
\begin{algorithmic}[1]
    \Function{$\leq_\mathds{A}$}{$A$, $A'$} \Comment{Compare, call written in infix notation, \textit{e.g.}, $A \leq_\mathds{A}  A'$}
        \State $\textit{created} \leftarrow A_\uparrow \leq A'_\uparrow$
        \State $\textit{burned} \leftarrow A_\downarrow \leq A'_\downarrow$
        \State $\textit{given} \leftarrow A_{\rightarrow *} \subseteq A'_{\rightarrow *} \wedge \bigwedge\limits_{id \in A_{\rightarrow *}} A_{\rightarrow}[id] \leq A'_{\rightarrow}[id]$ \Comment{Keys of $A_\rightarrow$ written $A_{\rightarrow *}$}
        \State $\textit{acked} \leftarrow A_{\leftarrow *} \subseteq A'_{\leftarrow *} \wedge \bigwedge\limits_{id \in A_{\leftarrow *}} A_{\leftarrow}[id] \leq A'_{\leftarrow}[id]$
        \State \Return $A_{\footnotesize \textbf{id}} = A'_{\footnotesize \textbf{id}} \wedge \textit{created} \wedge \textit{burned} \wedge \textit{given} \wedge \textit{acked}$
    \EndFunction
    \State
    \Function{$\sqcup_\mathds{A}$}{$A$, $A'$} \Comment{Merge, call written in infix notation, \textit{e.g.}, $A \sqcup_\mathds{A}  A'$}
    	\If{$A_{\footnotesize \textbf{id}} \neq A'_{\footnotesize \textbf{id}}$}
		\textbf{error}
	\EndIf
	\State
        \State $A'' \leftarrow \texttt{initialize}(A_{\footnotesize \textbf{id}})$
        \State $A''_\uparrow \leftarrow \texttt{max}(A_\uparrow, A'_\uparrow)$
        \State $A''_\downarrow \leftarrow \texttt{max}(A_\downarrow, A'_\downarrow)$ 
        \State
        \State $R \leftarrow A_{\rightarrow *} \cup A_{\rightarrow *}'$ \Comment{Receiver ids}
	\For{$id ~\textbf{in}~ R$}
		\If{$id \in A_{\rightarrow *} \wedge id \in A'_{\rightarrow *} $}
			\State $A''_{\rightarrow}[id] \leftarrow \texttt{max}(A_{\rightarrow}[id], A'_{\rightarrow}[id])$
		\ElsIf{$id \in A_{\rightarrow *}$}
			\State $A''_{\rightarrow}[id] \leftarrow A_{\rightarrow}[id]$
		\Else
			\State $A''_{\rightarrow}[id] \leftarrow A'_{\rightarrow}[id]$
		\EndIf
	\EndFor
	\State
	\State $S \leftarrow A_{\leftarrow *} \cup A_{\leftarrow *}'$ \Comment{Sender ids}
	\For{$id ~\textbf{in}~S$}
		\If{$id \in A_{\leftarrow *} \wedge id \in A_{\leftarrow *}'$}
			\State $A''_{\leftarrow}[id] \leftarrow \texttt{max}(A_{\leftarrow}[id], A'_{\leftarrow}[id])$
		\ElsIf{$id \in A_{\leftarrow *}$}
			\State $A''_{\leftarrow}[id] \leftarrow A_{\leftarrow}[id]$
		\Else
			\State $A''_{\leftarrow}[id] \leftarrow A'_{\leftarrow}[id]$
		\EndIf
	\EndFor
	\State
	\State \Return $A''$	
    \EndFunction
\end{algorithmic}
\caption{\label{alg:account-ordering} Account: Ordering and Merging}
\end{algorithm}

The account merge operation, $\sqcup_\mathds{A}$, (Alg.~\ref{alg:account-ordering}) is also a monotonic operation that combines different states for the same account, possibly resulting from concurrent operations. It computes the smallest state $A''$ from any two states $A$ and $A'$, such that $A''$ is equivalent to having applied the operations that led to both $A$ and $A'$. In mathematical terms, $A''$ is the \textit{least upper bound} of both $A$ and $A'$ according to $\leq_\mathds{A}$. Because it is a least upper bound, merging two states $A$ and $A'$ such that $A$ has happened before $A'$, \textit{i.e.} $A \leq_\mathds{A} A'$, results in a new state $A'' = A'$. This property allows merging with duplicate and out-of-order previous states without impacting the result. Moreover, because a later merged state implicitly contains the updates that led to previous states across replicas, the convergence of all replicas to the same state only requires the latest state of all replicas to have been propagated to other replicas \textit{transitively}, \textit{i.e.} through intermediate replicas. This requires less messages and better tolerates lost messages than operation-based approaches, which require explicit broadcast of every operation to all replicas~\cite{shapiro:hal-00932836}. The merge operation is implemented by merging all counters from both states and taking the maximum value of those that are present on both states.\footnote{This implementation is not strictly equivalent to merging all operations that have happened across replicas because taking the maximum value of two resulting counter is not equivalent to the sum of increments from their previous common state: \textit{i.e.} if $c' = x + c$ and $c'' = y + c$, $max(c', c'')$ is equal to $c + x + y$ only if $x$ or $y$ is $0$. In effect, the merge effectively considers two concurrent increments as having a shared component that is not duplicated even if done concurrently. This subtle distinction is what enables not having to reason about ordering between operation updates. Concurrent operations on different counters are however equivalent to merging all operations.}

The state and monotonic operations  of Section~\ref{sec:account:state-operations} augmented with the partial ordering induced by relation $\leq_\mathds{A}$ and the account merge operation $\sqcup_\mathds{A}$ is a Conflict-Free Replicated Data Type~\cite{shapiro:hal-00932836} (with proof in Section~\ref{sec:proof:account}) because all replicas are guaranteed to converge to the same state eventually. However, concurrent updates may result in a negative balance: \textit{e.g.}, two concurrent burn and give operations may have a sufficient balance to complete independently and concurrently but the combined net previous balance may not be sufficient to cover both.\footnote{As follows from the previous note, if there were no transfers, concurrent burns would never result in a negative balance because only the largest burnt amount, which was locally valid, would be retained instead of the sum of all burnt amounts.} That being said, \textit{only concurrent updates} may result in a negative balance: if all updates are sequential, the pre-condition on crediting operations (Alg.~\ref{alg:account}, line~\ref{ln:burn-balance-check} and \ref{ln:give-balance-check}) prevents the balance from going negative (see proof in Section~\ref{section:proof:sequential-non-negative}).

\subsubsection{Negative Balance Condition}

While not necessary to define a state-based CRDT, we define an additional operation $\sqcap_\mathds{A}$ (Alg.~\ref{alg:account-hlb}) that is the symmetrical opposite of the merge operation: \textit{i.e.}, the \textit{greatest lower bound} of two account states. This definition completes the semi-lattice formed by the $\leq_\mathds{A}$ relation and the $\sqcup_\mathds{A}$ operation into a complete lattice. Informally, $A'' = A \sqcap_\mathds{A} A'$ computes the resulting state $A''$ from the largest set of updates that led to both $A$ and $A'$. It is implemented by keeping the minimum value of counters that are present in both states.

\begin{algorithm}
\begin{algorithmic}[1]
    \Function{$\sqcap_\mathds{A}$}{$A$, $A'$} \Comment{Call written in infix notation, \textit{e.g.}, $A \sqcap_\mathds{A}  A'$}
    	\If{$A_{\footnotesize \textbf{id}} \neq A'_{\footnotesize \textbf{id}}$}
		\textbf{error}
	\EndIf
	\State
        \State $A'' \leftarrow \texttt{initialize}(A_{\footnotesize \textbf{id}})$
        \State $A''_\uparrow \leftarrow \texttt{min}(A_\uparrow, A'_\uparrow)$
        \State $A''_\downarrow \leftarrow \texttt{min}(A_\downarrow, A'_\downarrow)$ 
	\For{$id ~\textbf{in}~ A_{\rightarrow *} \cap A_{\rightarrow *}'$} \Comment{Receiver ids}
		\State $A''_{\rightarrow}[id] \leftarrow \texttt{min}(A_{\rightarrow}[id], A'_{\rightarrow}[id])$
	\EndFor
	\For{$id ~\textbf{in}~ A_{\leftarrow *} \cap A_{\leftarrow *}'$} \Comment{Sender ids}
		\State $A''_{\leftarrow}[id] \leftarrow \texttt{min}(A_{\leftarrow}[id], A'_{\leftarrow}[id])$
	\EndFor
	\State \Return $A''$	
    \EndFunction
\end{algorithmic}
\caption{\label{alg:account-hlb} Account: Greatest Lower Bound}
\end{algorithm}

We use $\sqcap_\mathds{A}$ to establish the precise conditions under which merging two account states with non-negative balances may result in a new state with a negative balance.  Informally, there must have been more credits issued than debits received from the non-overlapping updates implicitly contained in $A$ and $A'$, as well as not enough reserves to cover the difference. Formally, the balance of the merge result of $A$ and $A'$, \textit{i.e.}, \texttt{balance}($A \sqcup_\mathds{A} A'$), will be negative if and only if the sum of differences in credits between $A$ and $A'$ is larger than the sum of differences in debits between $A$ and $A'$, and \texttt{balance}($A \sqcap_\mathds{A} A'$) was sufficiently low (see proof in Section~\ref{sec:proofs:merge-concurrent-accounts}):

\begin{equation}
\texttt{balance}(A \sqcup_\mathds{A} A') < 0 \Leftrightarrow  \Delta_\downarrow  + \Delta_\rightarrow >  \Delta_\uparrow  + \Delta_\leftarrow + \texttt{balance}(A \sqcap_\mathds{A} A')
\end{equation}

This completes our presentation of a single account CRDT. The next section explains how to combine multiple accounts.

\subsection{Ledger}
\label{sec:ledger}

Ledger replicas track the most up-to-date state of a set of accounts. They are essentially implemented as a grow-only dictionary of account replicas (Alg.~\ref{alg:ledger}), so the implementation of state and operations is straight-forward.

\begin{algorithm}
\begin{algorithmic}[1]
   \Function{initialize}{}
    	\State $L \leftarrow \{ \}$ \Comment{Dictionary of $\textit{id} \rightarrow \textit{accounts}$ }
    	\State \Return $L$
    \EndFunction
    \State
    \Function{add}{L, A}
        \State $L' \leftarrow \textit{copy}(L)$ \Comment{Deep copy}
    	\If{$A_{\scriptsize\textbf{id}} \notin L_*$}
		\State $L'[A_{\scriptsize\textbf{id}}] \leftarrow A$
	\Else
		\State $L'[A_{\scriptsize\textbf{id}}] \leftarrow L[A_{\scriptsize\textbf{id}}] \sqcup_A A$ \Comment{$\sqcup_A$ definition in Alg.~\ref{alg:account-ordering}}
	\EndIf
	\State \Return $L'$
    \EndFunction
    \State
    \Function{$\leq_\mathds{L}$}{$L$, $L'$} 
    	\State \Return $L_* \subseteq L'_* \wedge \bigwedge_{id \in L_*} L[id] \leq_A L'[id]$ \Comment{$\leq_A$ definition in Alg.~\ref{alg:account-ordering}}
    \EndFunction
    \State
    \Function{$\sqcup_\mathds{L}$}{$L$, $L'$}
        \State $L'' \leftarrow \texttt{initialize}()$
        \State $I \leftarrow L_* \cup L'_*$
	\For{$id ~\textbf{in}~ I$}
		\If{$id \in L_* \wedge id \in L'_*$}
			\State $L''[id] \leftarrow  L[id] \sqcup_A L'[id]$  \Comment{$\sqcup_A$ definition in Alg.~\ref{alg:account-ordering}}
		\ElsIf{$id \in L_*$}
			\State $L''[id] \leftarrow L[id]$
		\Else
			\State $L''[id] \leftarrow L'[id]$	
		\EndIf
	\EndFor
	\State \Return $L''$	
    \EndFunction
    \State
    \Function{balances}{L}
        \State \Return $\{~ (\textit{id}, \texttt{balance}(L[id])) \textbf{~for~} id ~\textbf{in}~ L_* ~\}$  \Comment{$L_*$ returns the set of ids in $L$}
    \EndFunction
\end{algorithmic}
\caption{\label{alg:ledger} Ledger}
\end{algorithm}

The \texttt{initialize} operation creates a new dictionary $L$ representing the ledger. The \texttt{balances} operation returns the balance of account replicas stored in $L$, in a new dictionary also indexed by account identifiers. 

The \texttt{add} operation adds a new account replica in $L$, returning a new dictionary $L'$. If account $A$ is not already present in $L$, it creates a new entry with a key corresponding to $A$'s identifier and the state of $A$ as a value. Otherwise, it merges the state of account $A$ with the state of the stored replica in $L$ with the same identifier, updating the ledger in the process.

The compare relation for ledger states $L$, and $L'$, written $L \leq_L L'$, returns true if the state of $L$ is smaller or equal than the state of $L'$. It does not modify the state of either $L$ or $L'$. The comparison is true if and only if $L$'s keys are a subset of keys in $L'$ and every account associated to the keys ($id$s) in $L$ is smaller than the same account in $L'$, using the compare relation for accounts ($\leq_A$).

The merge operation between ledger states $L$ and $L'$, written $L'' = L \sqcup_L L'$, returns a new ledger state $L''$ that is the smallest state according to $\leq_L$ that is also larger than both $L$ and $L'$. It is implemented by taking 1) the union of all keys (identifiers) on both $L$ and $L'$ and 2) associating each keys to the most up-to-date corresponding account state. The second is implemented with the merge operation on accounts ($\sqcup_A$, Alg.~\ref{alg:account-ordering}) if the account is present in both states, otherwise it takes the state of the account either in $L$ or $L'$, whichever is present.

This ledger design is also a Conflict-Free Replicated Data Type (see proof in Section~\ref{sec:proof:ledger}) because all replicas are guaranteed to converge to the same state eventually. To the exception of the possibility that accounts may have negative balance, there is no additional issue related to concurrency: any concurrent account updates will result in a correct ledger. We outline the exact safety and liveness conditions on balances next in Section~\ref{sec:safety-liveness}, after a brief discussion on compatible system models.

\subsection{Compatible System Models}

The design we presented is compatible with crash recoveries~\cite{cachin2011introduction}, as long as replicas eventually recover and stay available long enough to exchange the latest state information. In case a replica crashes and never recovers, only the latest operations that could not be replicated prior to the crash will be lost. 

Our design is however not compatible with adversarial replicas because the latter could allow accounts that are not part of the creator set to create an arbitrary number of tokens, as well as transfer arbitrary amounts of tokens to other accounts regardless of their local balance.

\subsection{Safety and Liveness Properties of Account Balances}
\label{sec:safety-liveness}

We now outline the balance safety and liveness conditions of our design.

Given a set $\mathcal{A}$ of the largest state of every account (differentiated by $id$) across replicas and a set $\mathcal{C \subseteq \mathcal{A}}$ of accounts allowed to create tokens, the tokens circulating between accounts with non-negative balances either were created or were overspent but not burnt. Formally, the following equality always hold and is therefore a safety property of the design (see proof in Section~\ref{sec:proof:balance-safety}):

\begin{equation}
\label{eq:balance-safety}
\sum\limits_{A \in \mathcal{A}^{\geq 0}} \texttt{balance}(A) \leq  
		\sum\limits_{C \in \mathcal{C}} C_\uparrow 
		- \sum\limits_{A \in \mathcal{A}} A_\downarrow 
		+ \sum\limits_{A \in \mathcal{A}^{< 0}}  -\texttt{balance}(A)
\end{equation}

For deployments in which negative balances are ruled out by design, \textit{e.g.}, if concurrent operations on the same account are not possible, then the inequality simply says that tokens in circulation were created but not burnt.

A slight variation replacing $\mathcal{A}$ by a ledger $L$ such that $\texttt{values}(L) \subseteq \mathcal{A}$, locally holds as long as $L$ is a \textit{transitive closure} over all accounts, \textit{i.e.},  as long as all accounts that sent tokens that were acknowledged by accounts in $L$ are also in $L$. Formally, the inequality is true if $\texttt{values}(L)$ contains all accounts that are referred by $id$ in $A_{\leftarrow *}$ for every account $A$ in $\texttt{values}(L)$:

\begin{eqnarray*}
\label{eq:balance-safety-ledger}
 & \texttt{values}(L) = \{ A \in \mathcal{A} : \exists A' \in \texttt{values}(L) \wedge A_{\scriptsize \textbf{id}} \in A'_{\leftarrow *} \} 
\Rightarrow \\
& \sum\limits_{A \in \texttt{values}(L)^{\geq 0}} \texttt{balance}(A) \leq  
		\sum\limits_{C \in \mathcal{C}} C_\uparrow 
		- \sum\limits_{A \in \texttt{values}(L)} A_\downarrow 
		+ \sum\limits_{A \in \texttt{values}(L)^{< 0}}  -\texttt{balance}(A)
\end{eqnarray*}

This is the case because whatever the state of sender accounts, they cannot influence the balance of other accounts unless the tokens they sent were first acknowledged. If $L$ is not a transitive closure of accounts, the inequality may or may not hold: \textit{e.g.}, if $\mathcal{A}$ contains only two accounts $A$ and $C$ with different ids such that $C \in \mathcal{C}$, $A \notin \mathcal{C}$, $\texttt{values}(L) = \{ A \}$, and $C$ has given tokens to $A$, then it does not hold.

The inequality still holds for account states in the ledger that are not the latest among all replicas, as long as the state of every sender accounts $S$ in $L$ is greater or equal than they were at the time the receiver account $R$ acknowledged receiving the token. This is true if and only if $R_{\leftarrow}[S_{\scriptsize \textbf{id}}] \leq S_{\rightarrow}[R_{\scriptsize \textbf{id}}]$. This is always the case in concurrent execution in which all operations we presented for ledgers and accounts terminate but may not hold after a partial replication, as would happen if the merge operation between two ledger states were interrupted.

Finally, all the previous inequalities are eventually turned into strict equalities if every receiver account eventually acknowledges the tokens that were sent (see proof in Section~\ref{sec:proof:balance-liveness}). Using an inequality as a safety property, instead of enforcing the equality after each update, allows an eventually consistent implementation of a ledger.

This completes the presentation of the Grow-Only Counters Ledger (GOC-Ledger). The design should be compatible with any eventually-consistent replicated database, such as Secure-Scuttlebutt~\cite{kermarrec2020gossiping}. The properties we claimed for the design are supported with proofs in the next section. The system costs of core operations and required supporting mechanisms, which would depend on the deployment environment, have not yet been characterized: they will be covered in future work. We would happy to hear about any 

\section{Proofs}
\label{sec:proofs}

In this section, we prove the properties discussed informally in Section~\ref{sec:design}. To make the proofs more accessible to practitioners and non-mathematicians, we follow the structured proof approach suggested by Lamport~\cite{lamport2012write}.

\subsection{Definitions}

We first make the semantics of algorithms more precise with the following definitions:
\begin{proof}
	\begin{pfenum}
		\item $\mathds{R}^+$ is the set of real positive numbers, including zero.
		\item $\mathds{I}$ is the set of all possible identifiers.
		\item $\mathds{C} \subseteq \mathds{I}$ is the set of identifiers allowed to create tokens.
		\item $\mathds{D}$ is the set of all possible \textit{grow-only dictionaries of grow-only counters} indexed by identifiers, \textit{i.e.}, $\mathds{D}$ is the power set of the cartesian product of $\mathds{I}$ and $\mathds{R}$ ($\mathds{D} = \mathcal{P}(\mathds{I} \times \mathds{R}))$, with the additional constraint that for any $D \in \mathds{D}$ and any $(\textit{id},r)$ such that $\textit{id} \in \mathds{I}$ and $r \in \mathds{R}$, $id$ appears in at most one tuple in $D$.
		\item $D \leq_\mathds{D} D' \Leftrightarrow D_* \subseteq D'_* \wedge \bigwedge\limits_{id \in D_*} D[id] \leq D'[id] $
		\item $\mathds{A}$ is the set of all possible account states, \textit{i.e.}, $\mathds{A}$ is the cartesian product $\mathds{I} \times \mathds{R}^+ \times \mathds{R}^+ \times \mathds{D} \times \mathds{D}$, such that $(A_{\footnotesize \textbf{id}}, A_\uparrow, A_\downarrow, A_\rightarrow, A_\leftarrow) \in \mathds{A}$. 
		\item $\mathds{L}$ is the set of all possible ledger states, \textit{i.e.} the power set of the cartesian product of $\mathds{I}$ and $\mathds{A}$ ($\mathds{L} = \mathcal{P}(\mathds{I} \times \mathds{A})$) with the additional constraints that for any $L \in \mathds{L}$ and any $(\textit{id}, A) \in L$ $\textit{id} = A_{\footnotesize \textbf{id}}$ and \textit{id} appears in at most one tuple in $L$.
	\end{pfenum}
\end{proof}

\subsection{Lemmas}

In this section, we prove one lemma that is used later to simplify a proof.

\subsubsection{The composition of least upper bounds is a least upper bound.}
\label{sec:lemma:lub-composition}
 
\begin{proof}
	\assume{\begin{pfenum}
		\item $\mathds{A}$ and $\mathds{B}$ are sets
		\item Binary relations $\leq_\mathds{A}$ and $\leq_\mathds{B}$ define semi-lattices $\mathds{S}_\mathds{A}$ and $\mathds{S}_\mathds{B}$
		\item Equality $=$ relationship exists for $\mathds{A}$ and $\mathds{B}$
		\item Binary operators $\alpha$ and $\beta$ compute least upper bounds, respectively in $\mathds{S}_\mathds{A}$ and $\mathds{S}_\mathds{B}$
	\end{pfenum}}
	\define{\begin{pfenum}
		\item $a <_\mathds{A} a' \defeq a \leq_\mathds{A} \wedge~ a \neq a'$ and $b <_\mathds{B} b' \defeq a \leq_\mathds{B} \wedge~ b \neq b'$
		\item $\mathds{C} = \mathds{A} \times \mathds{B}$, \textit{i.e.}, the cartesian product of $\mathds{A}$ and $\mathds{B}$ such that $C \in \mathds{C} \Rightarrow (C = (a,b)) \wedge a \in \mathds{A} \wedge b \in \mathds{B}$
		\item $C \leq_\mathds{C} C' \Leftrightarrow (a,b) \leq_\mathds{C} (a',b') \Leftrightarrow a \leq_\mathds{A} a' \wedge~ b \leq_\mathds{B} b'$ which defines a semi-lattice $\mathds{S}_\mathds{C}$
		\item $C = C'  \Leftrightarrow (a,b) = (a',b') \defeq a = a' \wedge b = b'$
		\item $C <_\mathds{C} C'  \defeq C \leq_\mathds{C} C' \wedge C \neq C'$
		\item $C ~\tau~ C' \Leftrightarrow (a,b)~\tau~(a',b') \defeq (a~\alpha~a', b~\beta~b')$
	\end{pfenum}}
	\prove{$C''  = C ~\tau~ C'$ is the least upper bound of $C$ and $C'$ in $\mathds{S}_\mathds{C}$}

	\step{lbl:lub:def-1}{$C \leq_\mathds{C} C''$}
	\begin{proof}
		\step{}{$C'' = C ~\tau~ C' = (a,b) ~\tau~ (a',b') = (a ~\alpha~ a', b ~\beta~ b') = (a'', b'')$}
		\begin{proof}
			By definition.
		\end{proof}
		
		\step{lbl:lub:def-1:1}{$a \leq_\mathds{A} (a ~\alpha~ a') = a''$}
		\begin{proof}
			By assumption on $\alpha$ and the definition of least upper bound.
		\end{proof}

		\step{lbl:lub:def-1:2}{$b \leq_\mathds{B} (b ~\beta~ b') = b''$}
		\begin{proof}
			By assumption on $\beta$ and the definition of least upper bound.
		\end{proof}
		
		\qedstep
		\begin{proof}
			By \stepref{lbl:lub:def-1:1} and \stepref{lbl:lub:def-1:2} and definition.
		\end{proof}
	\end{proof}
		
	\step{lbl:lub:def-2}{$C' \leq_\mathds{C} C''$}
	\begin{proof}
		\textit{Idem} \stepref{lbl:lub:def-1} by replacing $C$ for $C'$.
	\end{proof}	
		
	\step{lbl:lub:def-3}{$\nexists C''' \in \mathds{C} : C \leq_\mathds{C} C''' \wedge C' \leq_\mathds{C} C''' \wedge C''' <_\mathds{C} C'' $}
	\begin{proof}
		\step{lbl:lub:def-3-1}{
			\assume{\begin{pfenum}
				\item $a \in \mathds{A}$
				\item $a' \in \mathds{A}$
				\item $a'' = a ~\alpha~ a'$
			\end{pfenum}}
			\prove{$\nexists a''' \in \mathds{A} : a \leq_\mathds{A} a''' \wedge a' \leq_\mathds{A} a''' \wedge a''' <_\mathds{A} a'' $}
			
			\pf~By assumption, because $\alpha$ computes a least upper bound.
		}
		
		\step{lbl:lub:def-3-2}{
			\assume{\begin{pfenum}
				\item $b \in \mathds{B}$
				\item $b' \in \mathds{B}$
				\item $b'' = b ~\beta~ b'$
			\end{pfenum}}
			\prove{$\nexists b''' \in \mathds{B} : b \leq_\mathds{B} b''' \wedge b' \leq_\mathds{B} b''' \wedge b''' <_\mathds{B} b'' $}
			
			\pf~By assumption, because $\beta$ computes a least upper bound.
		}
		
		\qedstep
		\begin{proof}
			By the conjunction of \stepref{lbl:lub:def-3-1} and \stepref{lbl:lub:def-3-2}, associativity of $\wedge$ and definitions.
		\end{proof}
	\end{proof}
		
	\qedstep
	\begin{proof}
		By the conjunction of \stepref{lbl:lub:def-1}, \stepref{lbl:lub:def-2}, and \stepref{lbl:lub:def-3}, which is the definition of a least upper bound.
	\end{proof}
\end{proof}

\subsection{Convergence}

To establish the convergence of both accounts and ledger, we need to establish that their states form a \textit{monotonic semi-lattice}~\cite{shapiro:hal-00932836}, which involves three propositions: First, that all possible states can be organized in a semi-lattice $\mathds{S}$ ordered by $\leq$. This is a requisite for the next two properties. Second, that merging two states computes their \textit{Least Upper Bound} (LUB) in $\mathds{S}$. This ensures that the merge is \textit{commutative}, \textit{associative}, and \textit{idempotent}, providing \textit{safety}, \textit{i.e.}, that replicas will agree on the final state regardless of ordering, delays, or duplication of merge operations. Third, that all operations modify the state $S$ of a replica such that the new state $S'$ is either equal or larger than the previous state $S$ in $\mathds{S}$ (\textit{monotonicity}). This ensures all state changes will be eventually reflected in the new state of all replicas, either because the same update(s) will have concurrently been applied or because the new state will be the result of a merge. Assuming an underlying communication medium that ensures new states to be eventually delivered to other replicas, the three propositions combined ensure both \textit{liveness} and \textit{safety}: all state changes are going to be replicated on all replicas \textit{and} all replicas will agree on the final state automatically, i.e. \textit{strong eventual consistency}~\cite{shapiro:hal-00932836}.

Because an account state is the composition of \textit{grow-only counters} ($A_\uparrow$ and $A_\downarrow$) and \textit{grow-only dictionaries of grow-only counters}~\cite{lavoie2023statebased} ($A_\leftarrow$ and $A_\rightarrow$), and the ledger state is a \textit{grow-only dictionary of accounts}, the convergence proofs are straight-forward. We first prove that account replicas converge, then prove that ledgers converge as well.

\subsubsection{Account}
\label{sec:proof:account}

\begin{proof}
        \prove{The Account design listed in Alg.~\ref{alg:account}, ~\ref{alg:query}, and \ref{alg:account-ordering} is a state-based (convergent) CRDT.}
	\pfsketch ~An account is the composition of state-based grow-only counters and grow-only dictionaries of grow-only counters. The three properties of \textit{ordering}, \textit{least upper bound}, and \textit{monotonicity} that are sufficient to define a state-based CRDT are the conjunction of corresponding properties on grow-only counters and grow-only dictionaries of grow-only counters.
\end{proof}

\begin{proof}
	\step{lbl:account:ordering}{
		\textbf{Ordering:} 
		Ordering $\mathds{A}$ by $\leq_\mathds{A}$ (Alg.~\ref{alg:account-ordering}) forms a semi-lattice $\mathds{S}_\mathds{A}$.
	}
	\begin{proof}
		\pfsketch~ $\leq_\mathds{A}$ is the conjunction of the $\subseteq$ and $\leq$ relationships, respectively forming partial orders on sets of identifiers in dictionaries in $\mathds{D}$ and real positive numbers $\mathds{R}^+$ that compose the possible states. Specifically:
		\begin{pfenum}
			\item $\leq$ creates partial orders on $A_\downarrow$ and $A_\uparrow$ counters (\textit{created} and \textit{burned} booleans).
			\item conjunction of $\subseteq$ on dictionary keys and $\leq$ on dictionary values creates partial orders on $A_\rightarrow$ and $A_\leftarrow$ dictionaries, (\textit{given} and \textit{acked} booleans).
			\item The conjunction of \textit{created}, \textit{burned}, \textit{given}, and \textit{acked} is also a partial order.
		\end{pfenum} 		
		Detailed proofs have already been published for grow-only counters of natural numbers, grow-only dictionaries of grow-only counters, and the conjunction of partial orders~\cite{lavoie2023statebased}. Those proofs apply directly to this proposition by replacing natural numbers by real numbers and adding a conjunction of dictionaries and grow-only counters.
	\end{proof}

	\step{lbl:account:lub}{
	        \textbf{Least Upper Bound:} 
		\assume{\begin{pfenum}
			\item $A \in \mathds{A}, A' \in \mathds{A}$
			\item \label{lbl:account:assumption:id} $A_{\footnotesize \textbf{id}} = A'_{\footnotesize \textbf{id}}$
		\end{pfenum}}
		
		\prove{$A'' = A \sqcup_\mathds{A} A'$ is the LUB of $A$ and $A'$ in $\mathds{S}_\mathds{A}$.}}

		\begin{proof}
			\step{lbl:account:lub:counter}{ $A''_\uparrow = \texttt{max}(A_\uparrow, A'_\uparrow)$ is the least upper bound of $A_\uparrow$ and $A'_\uparrow$ in the semi-lattice of real numbers $\mathds{S}_{\mathds{R}^+}$ formed by $\leq$. Idem for $A''_\downarrow = \texttt{max}(A_\downarrow, A'_\downarrow)$ in $\mathds{S}_{\mathds{R}^+}$.}
			\begin{proof}
				Same proof as in Annex of~\cite{lavoie2023statebased}, but on real numbers instead.
			\end{proof}
			
			\step{lbl:account:lub:dict}{$A''_{\leftarrow}$ is the least upper bound of $A_{\leftarrow}$ and $A'_{\leftarrow}$  in the semi-lattice of dictionaries $\mathds{S}_\mathds{D}$ formed by $\leq_\mathds{D}$. Idem for $A''_{\rightarrow }$ in $\mathds{S}_\mathds{D}$ formed by $\leq_\mathds{D}$.
			}
			\begin{proof}
				Same proof as for dictionaries in Annex of~\cite{lavoie2023statebased}: their \texttt{compare} method on dictionaries is defined similarly as the  \textit{given} and \textit{acked} conditions of $\leq_\mathds{A}$ in Alg.~\ref{alg:account-ordering}.
			\end{proof}
			
			\qedstep
			\begin{proof}
				By \stepref{lbl:account:lub:counter} and \stepref{lbl:account:lub:dict} and LUB composition lemma of Section~\ref{sec:lemma:lub-composition}.
			\end{proof}
		\end{proof}

	\step{lbl:account:monotonicity}{
	\textbf{Monotonicity:}
	All operations that may generate a new state, when applied on account state $A$ and any possible arguments, result in a new account state either equal or larger than $A$ in $\mathds{S}_\mathds{A}$ according to $\leq_\mathds{A}$.}
	
	\begin{proof}
		\step{}{\case{$A = \texttt{Initialize}(id)$}}
		\begin{proof}
			Initializes a new $A$ but not from an existing state, so monotonicity does not apply.
		\end{proof}
		
		\step{}{\case{$A' = \texttt{create}(A,\textit{amount})$}}
		\begin{proof}
			\step{}{\case{$A_{\footnotesize \textbf{id}} \in \mathds{C} \wedge \textit{amount} > 0$}}
			\begin{proof}
				$A'_\uparrow = A_\uparrow + \textit{amount} > A_\uparrow$ and all others properties of $A'$ are equal to those of $A$, therefore $A <_\mathds{A} A'$.
			\end{proof}
			
			\step{}{\case{$A_{\footnotesize \textbf{id}} \notin \mathds{C} \vee \textit{amount} \leq 0$}}
			\begin{proof}
				$A=A'$
			\end{proof}
		\end{proof}
		
		\step{}{\case{$A' = \texttt{burn}(A,\textit{amount})$}}
		\begin{proof}
			\step{}{\case{$\textit{amount} > 0 \wedge \texttt{balance}(A) \geq \textit{amount}$}}
			\begin{proof}
				$A'_\downarrow = A_\downarrow + \textit{amount} > A_\downarrow$ and all others properties of $A'$ are equal to those of $A$, therefore $A <_\mathds{A} A'$. 
			\end{proof}
			
			\step{}{\case{$\textit{amount} \leq 0 \vee \texttt{balance}(A) < \textit{amount}$}}
			\begin{proof}
				$A=A'$
			\end{proof}
		\end{proof}
		
		\step{}{\case{$A' = \texttt{giveTo}(A,\textit{amount},\textit{id})$}}
		\begin{proof}
			\step{}{\case{$\textit{amount} > 0 \wedge \texttt{balance}(A) \geq \textit{amount}$}}
			\begin{proof}
				\step{}{\case{$\textit{id} \notin A_\rightarrow [id]$}}
				\begin{proof}
					$A_{\rightarrow *} \cup \{ \textit{id} \} = A'_{\rightarrow *}$ therefore $A_{\rightarrow *} \subset A'_{\rightarrow *}$. All other properties of $A'$ are equal to those of $A$, therefore $A <_\mathds{A} A'$.
				\end{proof}
				
				\step{}{\case{$\textit{id} \in A_\rightarrow[id]$}}
				\begin{proof}
					$A_\rightarrow[id] + \textit{amount} = A'_\rightarrow[id]$ therefore $A_\rightarrow[id] < A'_\rightarrow[id]$. All others properties of $A'$ are equal to those of $A$, therefore $A <_\mathds{A} A'$.
				\end{proof} 
			\end{proof}
			
			\step{}{\case{$\textit{amount} \leq 0 \vee \texttt{balance}(A) < \textit{amount}$}}
			\begin{proof}
				$A=A'$
			\end{proof}
		\end{proof}
		
		\step{}{\case{$A' = \texttt{ackFrom}(A,B)$}}
		\begin{proof}
			\step{}{\case{$A_{\scriptsize \textbf{id}} \in B_{\rightarrow *}$}}
			\begin{proof}
				\step{}{\case{$B_{\scriptsize \textbf{id}} \notin A_{\leftarrow *}$}}
				\begin{proof}
					$A_{\leftarrow *} \cup \{ B_{\scriptsize \textbf{id}} \} = A'_{\leftarrow *}$ therefore $A_{\leftarrow *} \subset A'_{\leftarrow *}$. All other properties of $A'$ are equal to those of $A$, therefore $A <_\mathds{A} A'$.
				\end{proof}
				
				\step{}{\case{$B_{\scriptsize \textbf{id}} \in A_{\leftarrow *}$}}
				\begin{proof}
					$A'_\leftarrow[B_{\scriptsize \textbf{id}}] = \texttt{max}(A_\leftarrow[B_{\scriptsize \textbf{id}}], B_\leftarrow[A_{\scriptsize \textbf{id}}])$ therefore $A_\leftarrow[B_{\scriptsize \textbf{id}}] \leq A'_\leftarrow[B_{\scriptsize \textbf{id}}]$. All others properties of $A'$ are equal to those of $A$, therefore $A \leq_\mathds{A} A'$.
				\end{proof} 
			\end{proof}
			
			\step{}{\case{$A_{\scriptsize \textbf{id}} \notin B_{\rightarrow *}$}}
			\begin{proof}
				$A=A'$
			\end{proof}
		\end{proof}
		
		\step{}{\case{$b = \texttt{balance}(A)$}}
		\begin{proof}
			Does not modify the state of $A$.
		\end{proof}
		
		\step{}{\case{$b = \texttt{unackedFrom}(A,B)$}}
		\begin{proof}
			Does not modify the state of $A$.
		\end{proof}
		
		\step{}{\case{$b = A \leq_\mathds{A} A'$}}
		\begin{proof}
			Does not modify the state of $A$ or $A'$.
		\end{proof}
		
		\step{}{\case{$A'' = A \sqcup_\mathds{A} A'$}}
		\begin{proof}
			$A \leq_\mathds{A} A'' \wedge A' \leq_\mathds{A} A''$ because $\sqcup_\mathds{A}$ computes a LUB in $\mathds{S}_\mathds{A}$.
		\end{proof}
	\end{proof}
	
	\qedstep
	\begin{proof}
		By the conjunction of \stepref{lbl:account:ordering}, \stepref{lbl:account:lub}, \stepref{lbl:account:monotonicity} which is the definition of a state-based CRDT.
	\end{proof}
\end{proof}

\subsubsection{Ledger}
\label{sec:proof:ledger}

\label{th:ledger-crdt}
\prove{The ledger of Algorithm~\ref{alg:ledger} is a state-based (convergent) conflict-free replicated data type.}
\begin{proof}
	\pfsketch~ The ledger is a grow-only dictionary of accounts, that are themselves state-based CRDTs.

\pf~		
	\step{lbl:ledger:ordering}{
		\textbf{Ordering:} 
		Ordering $\mathds{L}$ by $\leq_\mathds{L}$ (Alg.~\ref{alg:ledger}) forms a semi-lattice $\mathds{S}_\mathds{L}$.
	}
	\begin{proof}
		\pfsketch~ The ordering of $\mathds{L}$ is the conjunction the $\subseteq$ partial order on sets of identifiers that serve as dictionary keys, and $\leq_\mathds{A}$ on account states that serve as dictionary values. The proof is similar as for grow-only dictionaries of grow-only counters~\cite{lavoie2023statebased}.
	\end{proof}
	
	\step{lbl:ledger:lub}{
	        \textbf{Least Upper Bound:} 
		\assume{\begin{pfenum}
			\item $L \in \mathds{L}, L' \in \mathds{L}$
		\end{pfenum}}
		
		\prove{$L'' = L \sqcup_\mathds{L} L'$ is the LUB of $L$ and $L'$ in $\mathds{S}_\mathds{L}$.}}
	\begin{proof}		
		\step{lbl:ledger:lub:def-1}{$L \leq_\mathds{L} L''$}
		\begin{proof}
			\step{lbl:ledger:lub:def-1:1}{$L_* \subseteq L''_*$}
			\begin{proof}
				$L_* \subseteq (L_* \cup L'_*) = L''_*$.
			\end{proof}
			
			\step{lbl:ledger:lub:def-1:2}{$\forall_{\textit{id} \in L_*} L[\textit{id}] \leq_\mathds{A} L''[\textit{id}]$}
			\begin{proof}
				\step{lbl:ledger:lub:def-1:2:1}{\case{$\textit{id} \in L_* \wedge \textit{id} \in L'_*$}}
				\begin{proof}
					$ L[\textit{id}] \leq_\mathds{A} L[\textit{id}] \sqcup_\mathds{A}  L'[\textit{id}]  = L''[\textit{id}]$
				\end{proof}
				
				\step{lbl:ledger:lub:def-1:2:2}{\case{$\textit{id} \in L_* \wedge \textit{id} \notin L'_*$}}
				\begin{proof}
					$L[\textit{id}] = L''[\textit{id}]$
				\end{proof}
			\end{proof}
			
			\qedstep
			\begin{proof}
				$\leq_\mathds{L}$ is the conjunction of \stepref{lbl:ledger:lub:def-1:1} and \stepref{lbl:ledger:lub:def-1:2}.
			\end{proof}
		\end{proof}
		
		\step{lbl:ledger:lub:def-2}{$L' \leq_\mathds{L} L''$}
		\begin{proof}
			\textit{Idem} \stepref{lbl:ledger:lub:def-1} on $L'$ instead of $L$.
		\end{proof}
		
		\step{lbl:ledger:lub:def-3}{$\nexists L''' \in \mathds{L} : L \leq_\mathds{L} L''' \wedge L' \leq_\mathds{L} L''' \wedge L''' <_\mathds{L} L''$}
		\begin{proof}
			\pfsketch~ We show that both of the only two possible conditions which individually could make the proposition false are not possible, therefore the proposition is always true. 
			\step{lbl:ledger:lub:def-3:ids}{It is not possible for $L_* \subseteq L'''_* \wedge L'_* \subseteq L'''_* \wedge L'''_* \subset L''_*$. }
			\begin{proof}
				$L''_* = L_* \cup L'_*$ is the least upper-bound of $L_*$ and $L'_*$ in $\mathds{S}_{\mathcal{P}(\mathds{I})}$ ordered by $\subseteq$ because $\cup$ computes the least upper bound on sets of identifiers (see \cite{lavoie2023statebased}).  
			\end{proof}
			\step{lbl:ledger:lub:def-3:accounts}{
				\assume{$L'''_* \subseteq L''_*$ (otherwise it is not possible for $L''' <_\mathds{L} L''$)}
				\prove{$\forall_{\textit{id} \in L'''_*}$, it is not possible for $L'''[\textit{id}]  <_\mathds{A} L''[\textit{id}]$ and $L[\textit{id}] \leq_\mathds{A} L'''[\textit{id}]$ (if $id \in L_*$) and $L'[\textit{id}] \leq_\mathds{A} L'''[\textit{id}]$  (if $id \in L'_*$)} }
			\begin{proof}
				\step{}{\case{$\textit{id} \in L_* \wedge \textit{id} \notin L'_*$}}
				\begin{proof}
					$L''[\textit{id}] = L[\textit{id}]$
				\end{proof}
				
				\step{}{\case{$\textit{id} \notin L_* \wedge \textit{id} \in L'_*$}}
				\begin{proof}
					$L''[\textit{id}] = L'[\textit{id}]$
				\end{proof}
				
				\step{}{\case{$\textit{id} \in L_* \wedge \textit{id} \in L'_*$}}
				\begin{proof}
					$L''[\textit{id}] = L[\textit{id}] \sqcup_\mathds{A} L'[\textit{id}] = \textit{LUB}(L[\textit{id}], L'[\textit{id}])$
				\end{proof}
				
				\qedstep
				\begin{proof}
					Either $L''[id]$ is a least upper bound of $L[id]$ or $L''[id]$, or is equal to one or the other, therefore it is not possible.
				\end{proof}
			\end{proof}
			
			\qedstep
			\begin{proof}
				 \stepref{lbl:ledger:lub:def-3:ids} and \stepref{lbl:ledger:lub:def-3:accounts} are the only two possible conditions which individually could make the proposition false, and they are are not possible, therefore the proposition is always true. 
			\end{proof}
		\end{proof}

		\qedstep
		\begin{proof}
			The conjunction of \stepref{lbl:ledger:lub:def-1}, \stepref{lbl:ledger:lub:def-2}, \stepref{lbl:ledger:lub:def-3} is the definition of a least upper bound.
		\end{proof}
	\end{proof}

	\step{lbl:ledger:monotonicity}{		
	\textbf{Monotonicity:}
	All operations that may generate a new state, when applied on ledger state $L$ and any possible arguments, result in a new ledger state either equal or larger than $L$ in $\mathds{S}_\mathds{L}$ according to $\leq_\mathds{L}$.}
	\begin{proof}
		\step{}{\case{$L = \texttt{initialize}()$}}
		\begin{proof}
			Initializes a new $L$ but not from an existing state so monotonicity need not apply.
		\end{proof}
		
		\step{}{\case{$L' = \texttt{add}(L,A)$}}
		\begin{proof}
			\step{}{\case{$A_{\scriptsize \textbf{id}} \notin L_*$}}
			\begin{proof}
				$L_* \subset L_* \cup \{ A_{\scriptsize \textbf{id}} \} = L'_*$ and all dictionary values of $L$ are equal to those of $L'$.
			\end{proof}
			
			\step{}{\case{$A_{\scriptsize \textbf{id}} \in L_*$}}
			\begin{proof}
				$L_* = L'_*$, $L[A_{\scriptsize \textbf{id}}] \leq_\mathds{A} (L[A_{\scriptsize \textbf{id}}] \sqcup_\mathds{A} A) =  L'[A_{\scriptsize \textbf{id}}]$, and all other properties of $L'$ are equal to those of $L$.
			\end{proof}
		\end{proof}
			
		\step{}{\case{$b = L \leq_\mathds{L} L'$}}
		\begin{proof}
			Does not return a new ledger state so monotonicity need not apply.
		\end{proof}
			
		\step{}{\case{$L'' = L \sqcup_\mathds{L} L'$}}
		\begin{proof}
		 	By definition because $L''$ is a least upper bound (\stepref{lbl:ledger:lub}).
		\end{proof}
			
		\step{}{\case{$B = \texttt{balances}(L)$}}
		\begin{proof}
			Does not return a new ledger state so monotonicity need not apply.
		\end{proof}

	\end{proof}

	\qedstep
	\begin{proof}
		By the conjunction of \stepref{lbl:ledger:ordering}, \stepref{lbl:ledger:lub}, \stepref{lbl:ledger:monotonicity} which is the definition of a state-based CRDT.
	\end{proof}
\end{proof}

\subsection{Greatest Lower-Bound on Accounts}
\label{sec:hlb}

\assume{\begin{pfenum}
	\item $A \in \mathds{A}$
	\item $A' \in \mathds{A}$
	\item $A_{\scriptsize \textbf{id}} = A'_{\scriptsize \textbf{id}}$
\end{pfenum}}
\prove{\textbf{Greatest Lower Bound:} $A'' = A \cap_\mathds{A} A'$ (Algorithm~\ref{alg:account-hlb}) is the greatest lower bound of $A$ and $A'$ in $\mathds{S}_\mathds{A}$.}
\pfsketch~	The \texttt{minimum} function computes the greatest lower bound on numbers. The intersection ($\cap$) computes the greatest lower bound on sets of identifiers. The composition of greatest lower bounds is also a greatest lower bound on the cartesian products of domains (proof is similar to that for lowest upper bound of Section~\ref{sec:lemma:lub-composition}). $\cap_\mathds{A}$ is a composition of \texttt{min} on grow-only counters an $\cap$ on sets of identifiers, therefore it also computes an greatest lower bound.
\begin{proof}
	
\pf~	
Alternatively a direct proof, using the definition of greatest lower bound:
	\step{lbl:hlb-1}{$A'' \leq_\mathds{A} A$}
	\begin{proof}
		\step{lbl:hlb-1-1}{$A''_\downarrow = \texttt{min}(A_\downarrow, A'_\downarrow) \leq A_\downarrow$ \newline
			   $A''_\uparrow = \texttt{min}(A_\uparrow, A'_\uparrow) \leq A_\uparrow$
		}
				
		\step{lbl:hlb-1-2}{$A''_{\leftarrow *} = A_{\leftarrow *} \cap A'_{\leftarrow *} \subseteq A_{\leftarrow *}$ \newline
			   $A''_{\rightarrow *} = A_{\rightarrow *} \cap A'_{\rightarrow *} \subseteq A_{\rightarrow *}$
		}
		
		\step{lbl:hlb-1-3}{
			\assume{
				$id \in A_{\leftarrow *} \wedge id \in A'_{\leftarrow *}$
			}
			$A''_{\leftarrow}[id] = \texttt{min}(A_{\leftarrow}[id], A'_{\leftarrow}[id]) \leq A_{\leftarrow}[id]$
			
			\assume{
				$id \in A_{\rightarrow *} \wedge id \in A'_{\rightarrow *}$
			}
			$A''_{\rightarrow}[id] = \texttt{min}(A_{\rightarrow}[id], A'_{\rightarrow}[id]) \leq A_{\rightarrow}[id]$
		}
		
		\qedstep
		\begin{proof}
			The five required conditions for $\leq_\mathds{A}$ to be true (Alg.~\ref{alg:account-ordering}) are satisfied by the followings:
			\begin{pfenum}
				\item By assumption $A_{\scriptsize \textbf{id}} = A'_{\scriptsize \textbf{id}}$;
				\item ($\textit{created}$) By \stepref{lbl:hlb-1-1};
				\item ($\textit{burned}$) By \stepref{lbl:hlb-1-1};
				\item ($\textit{given}$) By \stepref{lbl:hlb-1-2} and \stepref{lbl:hlb-1-3} for all $id \in A''_{\rightarrow *}$, the required assumption on \stepref{lbl:hlb-1-3} all satisfied because $A''_{\rightarrow *} = A_{\rightarrow *} \cap A'_{\rightarrow *}$;
				\item ($\textit{acked}$) Idem but on $A''_{\leftarrow *}$.
			\end{pfenum}
		\end{proof}
	\end{proof}
	
	\step{lbl:hlb-2}{$A'' \leq_\mathds{A} A'$}
	\begin{proof}
		Idem \stepref{lbl:hlb-1} but on $A'$ instead of $A$.
	\end{proof}
	
	\step{lbl:hlb-3}{$\nexists A''' \in \mathds{A} : A''' \leq_\mathds{A} A \wedge A''' \leq_\mathds{A} A' \wedge A'' <_\mathds{A} A'''$}
	\begin{proof}
		Assume by contradiction that such a $A'''$ exists. Therefore one or multiple of these conditions should be true (taken from Alg.~\ref{alg:account-ordering} and removing the equality case):
		\begin{pfenum}
			\item $A''_\uparrow < A'''_\uparrow$
			\item $A''_\downarrow < A'''_\downarrow$
			\item $A''_{\leftarrow *} \subset A'''_{\leftarrow *}$
			\item $\forall id \in A''_{\leftarrow *} : A''_{\leftarrow}[id] < A'''_{\leftarrow}[id] $
			\item $A''_{\rightarrow *} \subset A'''_{\rightarrow *}$  
			\item $\forall id \in A''_{\rightarrow *} : A''_{\rightarrow}[id] < A'''_{\rightarrow}[id] $
		\end{pfenum}
		However, for every number comparison ($<$) a larger value is not smaller than the corresponding values of $A$ and $A'$ because it is larger than either or both. Similarly, for set comparisons ($\subset$) a larger subset is smaller than the corresponding identifiers sets of $A$ and $A'$ because it is larger than either or both.
		
		Since none of the conditions above can be true, such a $A'''$ does not exist.
	\end{proof}
	
	\qedstep
	\begin{proof}
		The conjunction of \stepref{lbl:hlb-1}, \stepref{lbl:hlb-2}, and \stepref{lbl:hlb-3} is the definition of a greatest lower bound.
	\end{proof}
\end{proof}

\subsection{Balance}
\label{sec:proof:balance}

In this section we present proofs related to the balance of accounts.

\subsubsection{Sequences of Account Operations Maintain a Non-Negative Balance}
\label{section:proof:sequential-non-negative}

\assume{\begin{pfenum}
	\item $A,B \in \mathds{A}$
	\item $\texttt{balance}(A) \geq 0$
	\item $x \in \mathds{R}$
	\item $\textit{id} \in \mathds{I}$
	\item $\mathds{C}$ is the set of identifiers allowed to create tokens
\end{pfenum}}
\prove{For any sequence of operations on $A$ resulting in a new state $A'$, $\texttt{balance}(A') \geq 0$.}
\begin{proof}
\pfsketch~ By induction.
	For every case that modifies the state of $A$, we list the properties that have changed between the original state of an account, and its following states (marked with $'$), that influence the balance computation. Every non-listed properties stays the same.
	
\pf~		
	\step{seq:nn-bal:create}{\case{$A' = \texttt{create}(A,x)$}}
	\begin{proof}
		\step{}{\case{$A_{\scriptsize \textbf{id}} \in \mathds{C} \wedge x > 0$}}
		\begin{proof}
			\begin{pfenum}
				\item $A'_\uparrow = A_\uparrow + x$
				\item $\texttt{balance}(A') = \texttt{balance}(A) + x > 0$
			\end{pfenum}
		\end{proof}
		
		\step{}{\case{$A_{\scriptsize \textbf{id}} \notin \mathds{C} \vee x \leq 0$}}
		\begin{proof}
			\begin{pfenum}
				\item $A'_\uparrow = A_\uparrow$
				\item $\texttt{balance}(A') = \texttt{balance}(A) \geq 0$
			\end{pfenum}
		\end{proof}		
	\end{proof}
	
	\step{seq:nn-bal:burn}{\case{$A' = \texttt{burn}(A,x)$}}
	\begin{proof}
		\step{}{\case{$x > 0 \wedge \texttt{balance}(A) \geq x$}}
		\begin{proof}
			\begin{pfenum}
				\item $A'_\downarrow = A_\downarrow + x$
				\item $\texttt{balance}(A') = \texttt{balance}(A) - x \geq 0$
			\end{pfenum}
		\end{proof}
		
		\step{}{\case{$x \leq 0 \vee \texttt{balance}(A) < x$}}
		\begin{proof}
			\begin{pfenum}
				\item $A'_\downarrow = A_\downarrow$
				\item $\texttt{balance}(A') = \texttt{balance}(A) \geq 0$
			\end{pfenum}
		\end{proof}		
	\end{proof}
	
	\step{seq:nn-bal:giveTo}{\case{$A' = \texttt{giveTo}(A,x,\textit{id})$}}
	\begin{proof}
		\step{}{\case{$x > 0 \wedge \texttt{balance}(A) \geq x$}}
		\begin{proof}
			\step{}{\case{$id \notin A_{\rightarrow *}$}}
			\begin{proof}
				\begin{pfenum}
					\item $A'_{\rightarrow *} = A_{\rightarrow *} \cup \{ \textit{id} \}$
					\item $A'_\rightarrow [\textit{id}] = x$
					\item $\texttt{balance}(A') = \texttt{balance}(A) - x \geq 0$
				\end{pfenum}
			\end{proof}
			
			\step{}{\case{$id \in A_{\rightarrow *}$}}
			\begin{proof}
				\begin{pfenum}
					\item $A'_\rightarrow [\textit{id}] = A_\rightarrow [\textit{id}] + x$
					\item $\texttt{balance}(A') = \texttt{balance}(A) - x \geq 0$
				\end{pfenum}
			\end{proof}
		\end{proof}
		
		\step{}{\case{$x \leq 0 \vee \texttt{balance}(A) < x$}}
		\begin{proof}
			\begin{pfenum}
				\item $A'_\downarrow = A_\downarrow$
				\item $\texttt{balance}(A') = \texttt{balance}(A) \geq 0$
			\end{pfenum}
		\end{proof}		
	\end{proof}
	
	\step{seq:nn-bal:ackFrom}{\case{$A' = \texttt{ackFrom}(A,B)$}}
	\begin{proof}
		\step{}{\case{$A_{\scriptsize\textbf{id}} \in B_{\rightarrow *} \wedge \texttt{balance}(B) \geq 0$}}
		\begin{proof}
			\step{}{\case{$B_{\scriptsize\textbf{id}}  \notin A_{\leftarrow *}$}}
			\begin{proof}
				\begin{pfenum}
					\item $A'_{\leftarrow *} = A_{\leftarrow *} \cup \{ \textit{id} \}$
					\item $A'_\leftarrow [\textit{id}] = B_{\rightarrow}[A_{\scriptsize\textbf{id}}]$
					\item $\texttt{balance}(A') = \texttt{balance}(A) +  B_{\rightarrow}[A_{\scriptsize\textbf{id}}] \geq 0$, because only \texttt{giveTo} may have modified $B_{\rightarrow}[A_{\scriptsize\textbf{id}}]$ and only by increasing by a positive number. 
				\end{pfenum}
			\end{proof}
			
			\step{}{\case{$B_{\scriptsize\textbf{id}}  \in A_{\leftarrow *}$}}
			\begin{proof}
				\begin{pfenum}
					\item $A'_\leftarrow [\textit{id}] = \texttt{max}(A_\leftarrow [\textit{id}], B_{\rightarrow}[A_{\scriptsize\textbf{id}}])$
					\item $\texttt{balance}(A') = \texttt{balance}(A) +  \texttt{max}(A_\leftarrow [\textit{id}], B_{\rightarrow}[A_{\scriptsize\textbf{id}}]) \geq 0$, for the same reason as above.
				\end{pfenum}
			\end{proof}

		\end{proof}
		
		\step{}{\case{$A_{\scriptsize\textbf{id}} \notin B_{\rightarrow *} \vee \texttt{balance}(B) < 0$}}
		\begin{proof}
			\begin{pfenum}
				\item $A' = A$
				\item $\texttt{balance}(A') = \texttt{balance}(A) \geq 0$
			\end{pfenum}
		\end{proof}		
	\end{proof}

	\qedstep
	\begin{proof}
		Since \stepref{seq:nn-bal:create}, \stepref{seq:nn-bal:burn}, \stepref{seq:nn-bal:giveTo}, and \stepref{seq:nn-bal:ackFrom} maintain a non-negative balance for any possible inputs, any sequence of those operations also maintains a non-negative balance.
	\end{proof}
\end{proof}

\subsubsection{Merging Non-negative Accounts May Result in a Negative Balance}
\label{sec:proofs:merge-concurrent-accounts}

It is somewhat obvious that if there is no restriction, giving the same tokens to different recipients (concurrently) may lead to a negative balance after merging the resulting states. However, it was not completely obvious how to precisely formulate this intuition in a state-based model. 

The first insight was to look at the difference between credits and debits in both states, which is expressed as $\Delta$ below. However, this was not sufficient because the final negative condition on the balance depends on another previous state: if the balance in the previous state is sufficiently high, even if more credits are made than debits received, the final balance would still be positive. 

The second and missing insight was that none of the states being compared can be used as the previous state: instead, one must extract the common part with the $\sqcap_\mathds{A}$ operator, which intuitively represents the longuest "overlapping" or common operations that led to both. This concepts covers both the largest state that actually happened before both $A$ and $A'$ as well as similar operations that happened independently on both $A$ and $A'$ that led to similar state changes. For example, consider the following execution: two replicas of the same account (same id) are initialized independently. Then 10 tokens are created on both independently, without merging after. Then on one replica 6 tokens are given to Bob resulting in $A$ and on the other replica 5 tokens are given to Alice resulting in $A'$. Even though the replicas never merged, there is still a shared part in the state which is $A \sqcap_\mathds{A} A'$ with the creation counter with a value of 10 tokens.

The power, but also the difficulty, in this formulation is that it abstract multiple kinds of histories that may lead to the same state. But it also connects to our regular intuition when $A \sqcap_\mathds{A} A'$ has a balance of 0, as is the case in the initial state: then the balance of the merged states will be negative if and only if there has been more credits issued than debits received. The proof itself follows easily from the two insights.

\define{\begin{pfenum}
	\item $A \in \mathds{A} : \texttt{balance}(A) \geq 0$
	\item $A' \in \mathds{A} : \texttt{balance}(A') \geq 0$
	
	\item $\Delta_\downarrow \defeq   |A_\downarrow - A'_\downarrow| $
	\item $\Delta_\rightarrow \defeq    \sum\limits_{id \in A_{\rightarrow *} \cap A'_{\rightarrow *}} |A_{\rightarrow}[id]-A'_{\rightarrow}[id]| + \sum\limits_{id \in A_{\rightarrow *} \backslash A'_{\rightarrow *}}A_{\rightarrow}[id]  +  \sum\limits_{id \in A'_{\rightarrow *} \backslash A_{\rightarrow *}}A'_{\rightarrow}[id] $
	\item $\Delta_\uparrow \defeq   |A_\uparrow - A'_\uparrow| $
	\item $\Delta_\leftarrow \defeq    \sum\limits_{id \in A_{\leftarrow *} \cap A'_{\leftarrow *}} |A_{\leftarrow}[id]-A'_{\leftarrow}[id]| + \sum\limits_{id \in A_{\leftarrow *} \backslash A'_{\leftarrow *}}A_{\leftarrow}[id]  +  \sum\limits_{id \in A'_{\leftarrow *} \backslash A_{\leftarrow *}}A'_{\leftarrow}[id] $
\end{pfenum}}
\assume{$A_{\scriptsize \textbf{id}} = A'_{\scriptsize \textbf{id}}$}
\prove{$\texttt{balance}(A \sqcup_\mathds{A} A') < 0 \Leftrightarrow  \Delta_\downarrow  + \Delta_\rightarrow >  \Delta_\uparrow  + \Delta_\leftarrow + \texttt{balance}(A \sqcap_\mathds{A} A')$}
\begin{proof}
\pfsketch~ The key idea is that the maximum between two counters is equal to the min of the two plus their absolute difference and the rest follows from algebric manipulations.
	
\pf~	
	\step{lbl:eq}{$\texttt{balance}(A \sqcup_\mathds{A} A') = \texttt{balance}(A \sqcap_\mathds{A} A') +  \Delta_\uparrow  + \Delta_\leftarrow  - \Delta_\downarrow  - \Delta_\rightarrow$}
	\begin{proof}
		\step{lbl:min-max-eq}{\assume{\begin{pfenum}
			\item $x,x' \in \mathds{R}$
		\end{pfenum}}
		\prove{$\texttt{max}(x,x') = \texttt{min}(x,x') + | x - x' |$}}
		\begin{proof}
			\step{lbl:3-1}{\case{$x > x'$}}
			\begin{proof}
				\step{}{$x = \texttt{min}(x,x') + | x - x' |$}
				\begin{proof}
					By assumption, therefore $\texttt{max}(x,x)' = x$.
				\end{proof}
				
				\step{}{$x = x' + | x - x' |$}
				\begin{proof}
					By assumption, therefore $\texttt{min}(x,x)' = x'$.
				\end{proof}
				
				\qedstep
				\begin{proof}
					Because $x - x' = | x - x' |$ and $x - x' > 0$.
				\end{proof}
			\end{proof}
			
			\step{lbl:3-2}{\case{$x < x'$}}
			\begin{proof}
				\step{}{$x' = \texttt{min}(x,x') + | x - x' |$}
				\begin{proof}
					By assumption, therefore $\texttt{max}(x,x)' = x'$.
				\end{proof}
				
				\step{}{$x' = x + | x - x' |$}
				\begin{proof}
					By assumption, therefore $\texttt{min}(x,x)' = x$.
				\end{proof}
				
				\qedstep
				\begin{proof}
					Because $x'-x = | x - x' |$ and $x' - x > 0$.
				\end{proof}
			\end{proof}
			
			\step{}{\case{$x = x'$}}
			\begin{proof}
				By replacing everything with $x$: $x = x + | x - x |$.
			\end{proof}
			
			\qedstep
			\begin{proof}
				No other cases and true for all of them.
			\end{proof}
		\end{proof}
		
		\step{}{\begin{pfenum}
			\item $\texttt{max}(A_\uparrow, A'_\uparrow) = \texttt{min}(A_\uparrow, A'_\uparrow) + | A_\uparrow - A'_\uparrow |$
			\item $\texttt{max}(A_\downarrow, A'_\downarrow) = \texttt{min}(A_\downarrow, A'_\downarrow) + | A_\downarrow - A'_\downarrow |$
		\end{pfenum}}
		\begin{proof}
			By replacing $x$ respectively with $A_\uparrow$ and $A_\downarrow$ in \stepref{lbl:min-max-eq}.
		\end{proof}
		
		\step{}{\assume{\begin{pfenum}
			\item $id \in A_{\leftarrow *} \wedge id \in A'_{\leftarrow *}$
		\end{pfenum}}
		\prove{\begin{pfenum}
			\item $\texttt{max}(A_{\leftarrow}[id],A'_{\leftarrow}[id]) = \texttt{min}(A_{\leftarrow}[id],A'_{\leftarrow}[id]) + | A_{\leftarrow}[id] - A'_{\leftarrow}[id] |$
			\item $\texttt{max}(A_{\rightarrow}[id],A'_{\rightarrow}[id]) = \texttt{min}(A_{\rightarrow}[id],A'_{\rightarrow}[id]) + | A_{\rightarrow}[id] - A'_{\rightarrow}[id] |$
		\end{pfenum}}}
		\begin{proof}
			By replacing $x$ respectively with $A_\leftarrow[id]$ and $A_\rightarrow[id]$ in \stepref{lbl:min-max-eq}.
		\end{proof}
		
		\step{}{\define{\begin{pfenum}
			\item $\textit{debits}_{A \sqcup_\mathds{A} A'} =
							\texttt{max}(A_\uparrow, A'_\uparrow) 
		                                         + \sum\limits_{id \in A_{\leftarrow *} \cap A'_{\leftarrow *}}\texttt{max}(A_{\leftarrow}[id], A'_{\leftarrow}[id]) 
		                                         \newline
		                                         + \sum\limits_{id \in A_{\leftarrow *} \backslash A'_{\leftarrow *}}A_{\leftarrow}[id] 
		                                         + \sum\limits_{id \in A'_{\leftarrow *} \backslash A_{\leftarrow *}}A'_{\leftarrow}[id] 
		                                         $
		         \item $\textit{credits}_{A \sqcup_\mathds{A} A'} =
							\texttt{max}(A_\downarrow, A'_\downarrow) 
		                                         + \sum\limits_{id \in A_{\rightarrow *} \cap A'_{\rightarrow *}}\texttt{max}(A_{\rightarrow}[id], A'_{\rightarrow}[id]) 
		                                         \newline
		                                         + \sum\limits_{id \in A_{\rightarrow *} \backslash A'_{\rightarrow *}}A_{\rightarrow}[id] 
		                                         + \sum\limits_{id \in A'_{\rightarrow *} \backslash A_{\rightarrow *}}A'_{\rightarrow}[id] 
		                                         $                             
			\end{pfenum}}
		\prove{
			$\texttt{balance}(A \sqcup_\mathds{A} A') = \textit{debits}_{A \sqcup_\mathds{A} A'}+ \textit{credits}_{A \sqcup_\mathds{A} A'}$
		}}
		\begin{proof}
			By definition of \texttt{balance} (Alg.~\ref{alg:account}) and $A \sqcup_\mathds{A} A'$ (Alg.~\ref{alg:account-ordering}).
		\end{proof}
		
		\step{}{\define{\begin{pfenum}
			\item $\textit{debits}_{A \sqcap_\mathds{A} A'} =
							\texttt{min}(A_\uparrow, A'_\uparrow) 
		                                         + \sum\limits_{id \in A_{\leftarrow *} \cap A'_{\leftarrow *}}\texttt{min}(A_{\leftarrow}[id], A'_{\leftarrow}[id])
		                                         $
		         \item $\textit{credits}_{A \sqcap_\mathds{A} A'} =
							\texttt{min}(A_\downarrow, A'_\downarrow) 
		                                         + \sum\limits_{id \in A_{\rightarrow *} \cap A'_{\rightarrow *}}\texttt{min}(A_{\rightarrow}[id], A'_{\rightarrow}[id]) 
		                                         $                             
			\end{pfenum}}
		\prove{
			$\texttt{balance}(A \sqcap_\mathds{A} A') = \textit{debits}_{A \sqcap_\mathds{A} A'}+ \textit{credits}_{A \sqcap_\mathds{A} A'}$
		}}
		\begin{proof}
			By definition of \texttt{balance} (Alg.~\ref{alg:account}) and $A \sqcap_\mathds{A} A'$ (Alg.~\ref{alg:account-hlb}).
		\end{proof}
		
		\qedstep
		\begin{proof}
			By the previous steps and rearranging terms according to the $\Delta$ definitions.
		\end{proof}                      
	\end{proof}

	\step{}{$\texttt{balance}(A \sqcup_\mathds{A} A') < 0 \Leftrightarrow  \texttt{balance}(A \sqcap_\mathds{A} A') + \Delta_\uparrow  + \Delta_\leftarrow - \Delta_\downarrow  - \Delta_\rightarrow < 0$}
	\begin{proof}
		By applying inequality to both sides of \stepref{lbl:eq}.
	\end{proof}
	
	\qedstep
	\begin{proof}
		By rearranging terms.	
	\end{proof}
\end{proof}

\subsubsection{Safety: Net Tokens Owned were Created or Overspent, and not Burnt}
\label{sec:proof:balance-safety}

This safety property, expressed as an inequality instead of as an equality, is a key ingredient in making the design eventually consistent because this way the transfer of tokens between a sender and receivers is decomposed into two actions that allow the states of the sender and receiver accounts to be modified independently. Remarkably, this is possible even though there is a causal dependency between sending and receiving, which is enforced by checking that the sender counter is larger than the corresponding and symmetric receiver counter is increased (up to the same amount).

Moreover, we have two major differences compared to other works: 1) we track the creation of tokens within the same account abstraction, therefore we do not only enforce that sent tokens are eventually equal to received tokens globally; and 2) we also show the effect of overspent tokens in the inequality, which is that they increase the supply.  Surprisingly to us initially, the second aspect does not appear unless the accounts are segregated between those with a positive balance and those with a negative balance because the sending and reception of overspent tokens cancel each other if we sum the balances of all accounts. We therefore segregated accounts according to their balance in the inequality so that the right-hand side would correspond to net supply in circulation and the left-hand side would correspond to the net amount of tokens owned.

In our definitions below, we take a global view and reason over the global latest state of every account, regardless of the state of the local ledger replicas. This is the most general and intuitive stance. More restricted local views as discussed in the previous section follow easily from this so we only give the proof for the global case.

\define{\begin{pfenum}
	\item $\mathcal{L}$ is the set of all ledger replicas at any time
	\item $L =  \bigsqcup_\mathds{L} \mathcal{L}$ is the least upper bound on the state of replicas
	\item $\mathcal{A} = \{ L[id] : id \in L\}$ is the set of account states
	\item $\mathcal{C} = \{  A \in \mathcal{A} : A_{\scriptsize \textbf{id} \in \mathds{C}} \}$ are the accounts allowed to create tokens
	\item $\mathcal{A}^{\geq 0} = \{ A \in \mathcal{A} : \texttt{balance}(A) \geq 0 \} $ are the accounts with positive balances
	\item $\mathcal{A}^{< 0} = \{ A \in \mathcal{A} : \texttt{balance}(A) < 0 \} $ are the accounts with negative balances
\end{pfenum}}
\prove{
	$\sum\limits_{A \in \mathcal{A}^{\geq 0}} \texttt{balance}(A) \leq  
		\sum\limits_{C \in \mathcal{C}} C_\uparrow 
		- \sum\limits_{A \in \mathcal{A}} A_\downarrow 
		+ \sum\limits_{A \in \mathcal{A}^{< 0}}  -\texttt{balance}(A) $.}
\begin{proof}
	
\pf~	
	\step{}{$\sum\limits_{A \in \mathcal{A}} \texttt{balance}(A) \leq  
		\sum\limits_{C \in \mathcal{C}} C_\uparrow 
		- \sum\limits_{A \in \mathcal{A}} A_\downarrow$}
	\begin{proof}
		By sending  $\sum\limits_{A \in \mathcal{A}^{< 0}}  -\texttt{balance}(A)$ on the other side of the inequality, removing the double negation, and combining in a single summation.
	\end{proof}
	
	\step{}{$\sum\limits_{A \in \mathcal{A}} A_\uparrow
	 	   +\sum\limits_{A \in \mathcal{A}} \sum\limits_{id \in A_{\leftarrow}} A_\leftarrow[id]
		   -\sum\limits_{A \in \mathcal{A}} A_\downarrow
		   -\sum\limits_{A \in \mathcal{A}} \sum\limits_{id \in A_{\rightarrow}} A_\rightarrow[id] 
	 \leq  
		\sum\limits_{C \in \mathcal{C}} C_\uparrow 
		- \sum\limits_{A \in \mathcal{A}} A_\downarrow$}
	\begin{proof}
		By replacing \texttt{balance} by its definition.
	\end{proof}
	
	\step{lbl:sending-receiving-ineq-sums}{$
	 	\sum\limits_{A \in \mathcal{A}} \sum\limits_{id \in A_{\leftarrow}} A_\leftarrow[id]
	 \leq  
		\sum\limits_{A \in \mathcal{A}} \sum\limits_{id \in A_{\rightarrow}} A_\rightarrow[id] 
	 $}
	\begin{proof}
		Because $A_\uparrow = 0$ for all $A \notin \mathcal{C}$ and removing from both sides of the inequality. By removing $\sum\limits_{A \in \mathcal{A}} A_\downarrow$ from both sides of the inequality
	\end{proof}

	\step{lbl:sending-receiving-ineq-item}{For every pair of accounts $R,S \in \mathcal{A}$, $R_{\leftarrow}[S_{\scriptsize \textbf{id}}] \leq S_{\rightarrow}[R_{\scriptsize \textbf{id}}]$ always.}
	\begin{proof}
		The only operation that modifies $R_{\leftarrow}$ is \texttt{ackFrom} and every update either uses the current value of $S_{\rightarrow}[R_{\scriptsize \textbf{id}}]$, or the greatest of the previous value of  $R_{\leftarrow}[S_{\scriptsize \textbf{id}}]$, which can only be a previous value of  $S_{\rightarrow}[R_{\scriptsize \textbf{id}}]$, and $S_{\rightarrow}[R_{\scriptsize \textbf{id}}]$.
	\end{proof}
	
	\qedstep
	\begin{proof}
		Because \stepref{lbl:sending-receiving-ineq-item} is true for all pairs and $R_{\leftarrow}[S_{\scriptsize \textbf{id}}]$ exists only if $S_{\rightarrow}[R_{\scriptsize \textbf{id}}]$, \stepref{lbl:sending-receiving-ineq-sums} is true, which connects to the proposition by doing the algebric manipulations above in reverse.
	\end{proof}
\end{proof}

\subsubsection{Liveness: Eventual Balance Reconciliation}
\label{sec:proof:balance-liveness}

\define{~Same definitions as previous section.}
\prove{If no more \texttt{giveTo} operation is performed and every receiver account $R \in \mathds{A}$ eventually acknowledges all amounts sent by every sender account $S \in \mathds{A}$ (including $R=S$) with \texttt{ackFrom}, then eventually: \newline 
    $\sum\limits_{A \in \mathcal{A}^{\geq 0}} \texttt{balance}(A) =  
		\sum\limits_{C \in \mathcal{C}} C_\uparrow 
		- \sum\limits_{A \in \mathcal{A}} A_\downarrow 
		+ \sum\limits_{A \in \mathcal{A}^{< 0}}  -\texttt{balance}(A) $.}
\pf~ If all receiver accounts $R \in \mathcal{A}$ eventually acknowledge all amounts sent from sender accounts $S \in \mathcal{A}$ to $R$, then eventually $R_{\leftarrow}[S_{\scriptsize \textbf{id}}] = S_{\rightarrow}[R_{\scriptsize \textbf{id}}]$.  This turns the inequality from the previous section into an equality.

\section{Related Work}
\label{sec:related-work}

To the best of our knowledge, we are the first to implement an eventually-consistent ledger as a state-based Conflict Free Replicated Data Type (CRDT), including the ability for a subset of participants to create tokens. In this section, we survey related work.

\subsection{Consensus-Based Replicated Ledgers}

At the time of writing, the most popular approach to implement replicated ledgers is a \textit{blockchain}, \textit{i.e.} a global append-only log of updates with a total order that is established by the use of a consensus algorithm. The two most popular blockchains are Bitcoin~\cite{nakamoto2008bitcoin} and Ethereum~\cite{buterin2014next}. The large body of published work around blockchains is covered in multiple surveys that cover the main concepts behind blockchains~\cite{weichao2018blockchain,kolb2020survey-blockchains-ethereum} as well as specific aspects such as security and privacy~\cite{zhang2020survey-security-privacy-blockchains}, networking~\cite{dotan2022survey-blockchain-networking}, and smart-contracts implementing replicated state machines~\cite{kolb2020survey-blockchains-ethereum}. To improve scalability, some newer designs use a directed acyclic graph instead of an append-only log~\cite{kolb2020survey-blockchains-ethereum} but still rely on consensus to order updates.

Our design departs from blockchains in multiple ways. First, we decouple the sending and reception of tokens in a transaction: \textit{i.e.}, the sender account balance is immediately decreased but the receiver account balance is only increased \textit{eventually} after having replicated the sender account's latest state and acknowledged the sending. This decoupling enables an eventually consistent design. Second, token creation is done by a subset of accounts instead of through a mining protocol, as in Bitcoin~\cite{nakamoto2008bitcoin} or Ethereum~\cite{buterin2014next}. This approach enables \textit{local crypto-tokens}~\cite{lavoie2022localcryptotokens}. Third, our design covers the possibility of concurrent updates to the same account, which may result in a negative balance, while blockchains sequentially order all transactions \textit{among all accounts} to prevent this possibility. For applications for which a temporary negative balance is not critical (see Section~\ref{sec:economics}), our design is simpler and more efficient. Otherwise, for applications in which negative balances should be prevented, more frugal approaches suffice, as we discuss in the next section. 

\subsection{Consensus-Free Replicated Ledgers}
\label{sec:consensus-free-ledgers}

The total ordering of blockchain transactions is stronger than necessary to prevent overspending: sequential ordering of a single account's operations (Section~\ref{section:proof:sequential-non-negative}) or even only outgoing transactions from that account~\cite{guerraoui2021consensus} suffices. This insight, which to our knowledge has first been leveraged in Gupta's Master thesis~\cite{gupta2016nonconsensusdft} in 2016, has since motivated many subsequent theoretical designs and practical implementations. 

To our knowledge, all published work on consensus-free ledgers rely on broadcast abstractions~\cite{collins2020broadcast-payment,baudet2020fastpay,sliwinski2020abc,guerraoui2021consensus,auvolat2021money,frey:hal-03346756,kuznetsov2021permissionless,cholvi2021bdso,georghiades2021needs} that require reliable delivery of all updates. Our state-based approach works with unreliable channels, with fewer messages, and with incomplete communication topology, as long as state updates are eventually transitively delivered to all replicas. Our state approach has the added benefit of automatically combining multiple outgoing transfers to the same receiver into a single state update. That being said, the previous work tolerates adversarial environments while the design we presented does not. It is still an open question whether equivalent state-based solutions exist.

The practical benefits of foregoing consensus, resulting in much higher throughput and significantly lower latency compared to conventional blockchains, are dramatically illustrated by Astro~\cite{collins2020broadcast-payment} and Fast-Pay~\cite{baudet2020fastpay}. Both designs use managed replicas to ensure client operations are sequentially ordered, and therefore prevent overspending, even if any number of clients or less than 1/3 of replicas are adversarial (Byzantine). It is still an open question of how to design practical payment systems that would work in open adversarial settings, without resorting to 2/3 of trusted replicas within a managed environment.

\subsection{Conflict-Free Replicated Data Types} 

Conflict-Free Replicated Data Types (CRDTs)~\cite{shapiro:hal-00932836} are replicated mutable objects that are designed to ensure converge to the same state \textit{eventually}, \textit{i.e.}, at some point in the future after updates have stopped, and \textit{automatically}, \textit{i.e.} using deterministic conflict-resolution rules in the presence of concurrent updates. An initial survey of useful data types~\cite{shapiro:inria-00555588} kickstarted the field, which has blossomed into over a hundred publications, as listed on a dedicated website~\cite{crdt-website}.

To our knowledge, there are no other published state-based conflict-free replicated ledgers, as listed on the aforementioned website~\cite{crdt-website}. That being said, the community exploring CRDTs seems to run parallel to the distributed computing community with slightly different, but ultimately connected, formalisms. From a theoretical standpoint, Viotti and  Vukoli\'{c} surveyed all major consistency models that have been proposed for operations on distributed objects up to 2016, including CRDTs,  and organized them in a partial order~\cite{viotti2016consistency-db-survey}. Their survey shows that CRDTs consistency model runs parallel to most others published. From a practical standpoint, the designs presented in \cite{guerraoui2021consensus} and \cite{collins2020broadcast-payment} can be considered operation-based CRDTs since their operations commute, \textit{i.e.}, the order in which transfers are received does not matter as long as they are delivered in their sending order. Nonetheless, state-based replicated object designs are uncommon and mostly explored by the CRDT community, as illustrated by the fact that other replicated ledgers we have found all use an operational approach, based on reliable broadcast (see Section~\ref{sec:consensus-free-ledgers}).

A natural refinement of our state-based conflict-free replicated ledger would be to reformulate updates as a $\delta$-CRDT~\cite{Almeida2018delta}, a state-based CRDT approach that reduces the amount of state required to be sent based on the state already possessed by other replicas, either known from a stateful connection with other replicas~\cite{Almeida2018delta} or derived from meta-data~\cite{vanderLinde06delta}. We leave this for future work.

\subsection{Peer-to-Peer Eventually-Consistent Replicated Databases}
\label{sec:p2p-db}

The design we presented started as an exploration to implement a replicated ledger for Secure-Scuttlebutt (SSB)~\cite{kermarrec2020gossiping,tarr2019ssb}, an eventually consistent database based on single-writer append-only logs combined with peer-to-peer reconciliation protocols. Hypercore~\cite{hypercore-website}, previously known as DAT~\cite{ogden2017dat,robinson2018dat} and now maintained by Holepunch~\cite{holepunch-website}, is another eventually consistent database based on single-writer append-only logs. The append-only logs provide a sequential ordering based on cryptographic hashes and signatures, which would be sufficient to maintain non-negative account balances.  Both designs predate the equivalent exclusive logs used by Astro~\cite{collins2020broadcast-payment} (mentioned in Section~\ref{sec:consensus-free-ledgers}) and were inspired by a reconciliation protocol~\cite{vanRenesse2008reconciliation} published by Amazon researchers in 2008. 

However, the reconciliation protocol of Secure-Scuttlebutt does not tolerate \textit{forks}, \textit{i.e.}, a concurrent message inserted at the same index in a log, potentially intentionally by a malicious participant in their own log, that turns the log into a tree. In the presence of forks, replicas will only accept updates on the branch they initially started replicating and forever reject all others, which may result in an irreconcilable partitioning of replicas.  Hypercore suffers from the same issue~\cite{hypercore-handling-conflicts,hypercore-split-resolution-dep}  as SSB. Both designs are therefore not sufficient to implement our replicated ledger (and are not eventually consistent!) in an adversarial environment. How to deal with forks of single-writer append-only logs in a peer-to-peer adversarial environment is still an open question.

\subsection{Economics}
\label{sec:economics}

In self-sovereign cryptocurrencies~\cite{shapiro2023sovereign}, Shapiro identifies general desirable economic properties of cryptocurrencies that can be issued and traded by anyone (individuals, communities, corporations, municipalities, banks, etc.) and provide a protocol that meets these properties. In complement, Lavoie and Tschudin~\cite{lavoie2022localcryptotokens} highlight that transactions costs in regular consensus-based blockchains, such as Bitcoin and Ethereum, are larger than many local economic applications can support, argue for eventual detection rather than prevention of over-spending, and identify potential applications. The design of our conflict-free replicated ledger is a step in the direction of cryptotokens that meet both, but would require additional mechanisms to be fault-tolerant in adversarial settings.

To that end, core economic assumptions made by blockchains Bitcoin~\cite{nakamoto2008bitcoin} and Ethereum~\cite{buterin2014next}, and commonly assumed by many system designers of alternative systems, are worth revisiting. For example, the total number of tokens an account can create with our design is not bounded, which enables local applications in which the value of the tokens derives from the trust in the issuer rather than their scarcity~\cite{lavoie2022localcryptotokens}. Moreover, overspending, which results in a negative account balance, may not be much of an issue in some local economic applications. For example, drawing against a negative balance (up to a certain bound) is already a core aspect of mutual credit systems~\cite{schraven2001mutual}. It is also common for trusted participants in local economies to provide credit to one another: we have ourselves committed (and fulfilled our promise!) to pay back later after realizing we did not have enough money on us to pay for a haircut; we also witnessed other regular clients of a convenience store in Saint-Sulpice (Switzerland) do the same! 

%This possibility has started been explored theoretically: Bezerra and Kuznetsov~\cite{bezerra2022tame}, bound the number of times the same tokens can be spent with verifiable evidence that the same tokens have been spent multiple times and suggest possible strategies once overspending has been detected.

\section{Conclusion and Future Work}
\label{sec:conclusion}

We presented GOC-Ledger, a replicated ledger as a state-based conflict-free replicated data type based on grow-only counters. Our design lowers requirements on communication channels and topology between replicas, eventually converges even in the presence of negative balances without having to maintain a causal history of operations, and non-negative balances may be enforced by auxiliary mechanisms that ensure individual account updates are sequential.

We plan to design complementary techniques for deployments in adversarial environments, implement the design on top of eventually-consistent replicated databases, and measure its performance on realistic traces.

\section{Acknowledgements}
\label{sec:acknowledgements}

I would like to thank Christian F. Tschudin for fostering a research environment allowing detours and playfulness in the process, as well as providing financial support for this work and feedback on earlier versions of this work. The idea of investigating the possibilities offered by allowing accounts to converge to a negative balance in some circumstances was first suggested by him. Once the prejudice against negative balances was abandoned, that opened the possibility of expressing accounts as state-based CRDTs and greatly simplified the design.

I would also like to thank Christian F. Tschudin for having initiated a CRDT seminar and invited me to co-teach it, which created the opportunity of learning how to formalize CRDT algorithms. I was initially planning to write a system paper on a similar design. However, after 6-7 design iterations, it became obvious that the design had subtleties that greatly benefitted from a more formal treatment to tease out and get right. The teaching of the CRDT seminar happened exactly at the right time to review fundamentals of distributed algorithms in general, and CRDTs in particular, to ramp up the skills to carry such work.

I would finally like to thank Osman Biçer for feedback on early version of this draft, especially regarding proof presentation.

\newpage

\bibliographystyle{plainurl}
\bibliography{main}

\newpage
\appendix

\section{Notation and Conventions}
\label{apdx:notation}

We use notations and conventions that are good graphic mnemonics for the concepts and make the algorithms easier to reason about in the proofs (Section~\ref{sec:proofs}). The semantics are:

\begin{itemize}
	\item \textbf{Variables} are written in \textit{italic}:
		\begin{itemize}
			\item  lower case when containing literal values, ex: $id$;
			\item upper case when containing an object with multiple fields, a dictionary with multiple key-value pairs, or a set with multiple elements. For example, $A$ for account object, $L$ for a ledger dictionary, and $S$ for a set;
		\end{itemize}
	\item An \textbf{object's field} is accessed using a subscript, ex: $A$'s identifier stored in field $\textbf{id}$ is accessed $A_{\scriptsize \textbf{id}}$;
	\item An empty \textbf{dictionary} is written $\{\}$, accessing a dictionary $L$'s value stored under key $id$ is written $L[id]$, accessing all the keys of $L$ is written $L_*$;
	\item \textbf{Assigning} a new value to a variable, a field, or a dictionary entry uses $\leftarrow$, ex: $id \leftarrow id'$, $A_{\scriptsize\textbf{id}} \leftarrow id$, $L[id] \leftarrow A$. Variables, fields, and dictionaries are mutable and can be modified in place;
	 \item A \textbf{key-value} pair for dictionaries is written $\textit{key} \rightarrow \textit{value}$;
	 \item We use "dictionary-comprehension", similar to Python, for inline initialization fo dictionares, ex: ${ \textit{key} \rightarrow \textit{value} ~\textbf{for}~ \textit{key} ~\textbf{in}~ K }$;
	\item \textbf{Different states} for replicas of objects or dictionaries are written with $'$ and $''$ using the same variable name, ex: $A, A', A''$. We represent output values of functions using variable names with $'$ or $''$ to show they are later states of the same replica;
	\item The  \textbf{flow of tokens} on an account is suggested by the direction of an arrow used as a field name: 
		\begin{itemize}
			\item $A_\uparrow$ returns the number created tokens, which increases the account balance without transfers;
			\item $A_\downarrow$ returns the number of burned tokens, which decreases the account balance without transfers;
			\item $A_{\leftarrow}$ is a dictionary and $A_{\leftarrow}[id]$ returns the number of tokens received from $id$ and therefore flowing \textit{into} account $A$ (the arrow is a subscript to distinguish from assignment);
			\item $A_{\rightarrow}$ is a dictionary and $A_{\rightarrow}[id]$ returns the number of tokens given to $id$ therefore flowing \textit{out of} account $A$;
		\end{itemize}
\end{itemize}

Apart from these, we use common mathematical and pseudo-code conventions: 
\begin{itemize}
	\item $x \in X$ is an element $x$ in a set $X$ and $x \notin X$ means $x$ is not in a set $X$;
	\item $X \subseteq Y$ means $X$ is a subset of $Y$ which may include up to all elements of $Y$;
	 \item $\sum$ is a summation;
	 \item $\sum\limits_{x \in X} x$ is the sum of all elements in $X$;
	 \item $\leq$ is smaller or equal;
	 \item $\bigwedge$ and '$\textbf{and}$' both represent a logical \textit{and};
	 \item $\bigwedge\limits_{x \in X} x$ is the logical 'and' between all elements in $X$;
	 \item $\textbf{for}~x ~\textbf{in}~ X~\textbf{do}$ iterates over all values in $X$ sequentially assigning them to $x$.
\end{itemize}

\end{document}